\documentclass[prd,preprint,tightenlines,showpacs,showkeys,preprintnumbers,nofootinbib,eqsecnum,superscriptaddress]{revtex4}

 \usepackage[dvips,final]{graphicx}
  \usepackage{amssymb}
   \usepackage{amsmath}
    \usepackage{amsfonts}
     \usepackage{bm}

\begin{document}

\title{Exclusive scalar \mbox{\boldmath $f_0(1500)$} meson production \\
for energy ranges available \\ at the GSI Facility for Antiproton 
and Ion Research (GSI-FAIR) \\ and at the Japan Proton Accelerator
Research Complex (J-PARC)}

\author{A.~Szczurek}
\email{antoni.szczurek@ifj.edu.pl}
\affiliation{Institute of Nuclear Physics PAN, PL-31-342 Cracow,
Poland} 
\affiliation{University of Rzesz\'ow, PL-35-959 Rzesz\'ow, Poland}

\author{P.~Lebiedowicz}
\email{piotr.lebiedowicz@ifj.edu.pl}
\affiliation{Institute of Nuclear Physics PAN, PL-31-342 Cracow,
Poland} 
\affiliation{University of Rzesz\'ow, PL-35-959 Rzesz\'ow, Poland}

\date{\today}

\begin{abstract}
We evaluate differential distributions for exclusive scalar
$f_0(1500)$ meson (glueball candidate) production for 
$p \bar p  \to N_1 N_2 f_0$ (FAIR@GSI)
and $p p \to p p f_0$ (J-PARC@Tokai). Both QCD diffractive, 
pion-pion meson exchange current (MEC) components as well as 
double-diffractive mechanism with intermediate pionic loop 
are calculated for the first time in the literature. 
The pion-pion component, which can be reliably calculated, 
dominates close to the threshold 
while the diffractive component may take over only for larger energies.
At the moment only upper limit for the QCD-diffractive component
can be obtained. 
The diffractive component is calculated based on 
two-gluon impact factors as well as in the framework 
of Khoze-Martin-Ryskin approach proposed for diffractive 
Higgs boson production.
Different unintegrated gluon distribution functions (UGDFs) from 
the literature are used.  
Rather large cross sections due to pion-pion fusion are predicted 
for PANDA energies, where the gluonic mechanism is shown 
to be negligible.
The production of $f_0(1500)$ close to threshold 
could limit the so-called $\pi NN$ form factor in the region of 
larger pion virtualities.
We discuss in detail the two-pion background to the production
of the $f_0(1500)$ meson. 
\end{abstract}

\pacs{12.38.-t, 12.39.Mk, 14.40.Cs}
\keywords{exclusive production, pion-pion fusion, diffractive mechanisms, differential sross sections}

\maketitle

\section{Introduction}
Many theoretical calculations, including lattice QCD,
predicted existence of glueballs (particles dominantly made of 
gluons) with masses $M >$ 1.5 GeV. 
No one of them was up to now unambiguously identified.
The nature of scalar mesons below 2 GeV is also not well 
understood. 
Lattice QCD approach with quenched quarks find a scalar 
gluonium (glueball) at approximately 1.6 GeV \cite{MVC}
\footnote{The approaches with dynamical quarks find 
relatively large mixing with $q \bar q$ states \cite{UKQCD}.}.
Also the analyses in the framework of chiral 
Lagrangians \cite{FF,AO08} indicate that $f_0(1500)$
is dominantly gluonium state. The QCD Sum Rules 
\cite{NNN,HJZ,KGV,HS01,OSH01,SHO04} suggest that the states at approximately 
1 GeV and 1.5-1.6 GeV are admixtures of gluonium
and $q \bar q$ states.
A recent analysis in the framework of Gaussian QCD 
Sum Rules \cite{HMS08}, which is well suited to 
$q \bar q$ - gluonium mixing, find that the states
at about 1 GeV ($f_0(980)$) and at about 1.4 GeV
are strongly mixed with the preference of the 
higher-mass state to have slightly larger gluonium
admixture.
Summarizing this discussion, it may be very difficult
to find a clear signal of gluonium.
Further studies of the scalar meson production in
several processes may shed more light on the quite 
complicated problem.

The lowest mass meson considered as a glueball candidate is 
a scalar $f_0(1500)$ \cite{AC96} discovered by the Crystall Barrel
Collaboration in proton-antiproton annihilation 
\cite{Crystall_Barrel_f0_1500}.
The branching fractions are consistent with the dominant glueball 
component \cite{anisovich}.
It was next observed by the WA102 Collaboration in central
production in proton-proton collisions in two-pion
 \cite{WA102_f0_1500_2pi} and four-pion \cite{WA102_f0_1500_4pi}
decay channels at $\sqrt{s} \approx$ 30 GeV 
\footnote{No absolute normalization of the corresponding 
experimental cross section was available. 
Only two-pion or four-pion invariant mass spectra were discussed.}. 
Close and Kirk \cite{CK} proposed a phenomenological model
of central exclusive $f_0(1500)$ production. In their language
the pomerons (transverse and longitudinal) are the effective
(phenomenological) degrees of freedom \cite{CS99}.
The Close-Kirk amplitude was parameterized as
\footnote{The $\phi'$ dependence applies in the meson rest frame (current-current c.m.)}
\begin{equation}
{\cal M}(t_1,t_2,\phi') = a_T \exp\left(\frac{b_T}{2}(t_1+t_2)\right) 
         + a_L \frac{\sqrt{t_1 t_2}}{\mu^2} 
          \exp\left(\frac{b_L}{2}(t_1+t_2)\right)\;\cos (\phi') \; .
\end{equation}
In their approach there is no explicit $f_0(1500)$-rapidity 
dependence of the corresponding amplitude.
Since the parameters were rather fitted to the not-normalized 
WA102 data \cite{WA102_f0_1500_2pi} no absolute normalization can 
be obtained within this approach. Furthermore the parameterization
is not giving energy dependence of the cross section,
so predictions for other (not-measured) energies are not possible.
In the present paper we will investigate rather 
a QCD-inspired approach.
It provides absolute normalization \footnote{As will be discussed later it
is rather upper limit which can be easily obtained.}, energy dependence
and dependence on meson rapidity (or equivalently on 
$x_F$ of the meson).

The nature of the $f_0(1500)$ meson still remains rather 
unclear.
New large-scale devices being completed (J-PARC at Tokai) or 
planned in the future (FAIR at GSI) may open a new possibility to study
the production of $f_0(1500)$ in more details.

In the present analysis we shall concentrate on exclusive
production of scalar $f_0(1500)$ in the following reactions:
\begin{equation}
\begin{split}
p + p &\to p + f_0(1500) + p \; , \\
p + \bar p &\to p + f_0(1500) + \bar p \; , \\
p + \bar p &\to n + f_0(1500) + \bar n \; . \\
\end{split}
\end{equation}
While the first process can be measured at J-PARC, the latter
two reactions could be measured by the PANDA Collaboration
at the new complex FAIR planned in GSI Darmstadt.
The combination of these processes could shed more light
on the mechanism of $f_0(1500)$ production as well as 
on its nature.

\begin{figure}[!htb]    
 \centerline{\includegraphics[width=0.4\textwidth]{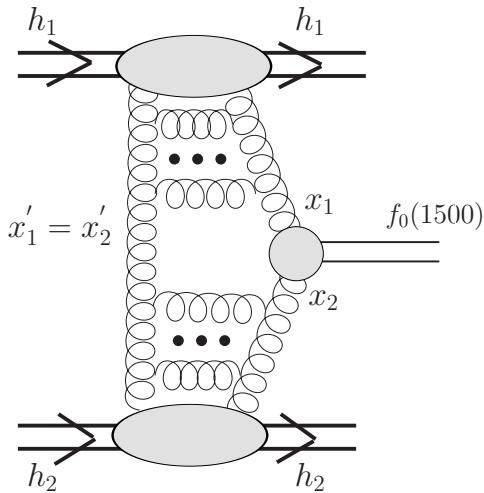}}
   \caption{\label{fig:QCDdiff}
   \small  The sketch of the bare QCD mechanism for diffractive
production of the glueball. The kinematical
variables are shown in addition.}
\end{figure}

If $f_0(1500)$ is a glueball (or has a strong glueball component \cite{CZ05})
then the mechanism shown in Fig. \ref{fig:QCDdiff}
may be important, at least in the high-energy regime.
This mechanism is often considered as the dominant mechanism of 
exclusive Higgs boson \cite{KMR}
and $\chi_c(0^+)$ meson \cite{PST07} production at high energies.
There is a hope to measure these processes at LHC in some future
when forward detectors will be completed.
At intermediate energies the same mechanism is, however, not able to
explain large cross section for exclusive $\eta'$ production 
\cite{SPT07} as measured by the WA102 Collaboration. 
Explanation of this fact is not clear to us in the moment.

\begin{figure}[!htb]    
 \centerline{\includegraphics[width=0.3\textwidth]{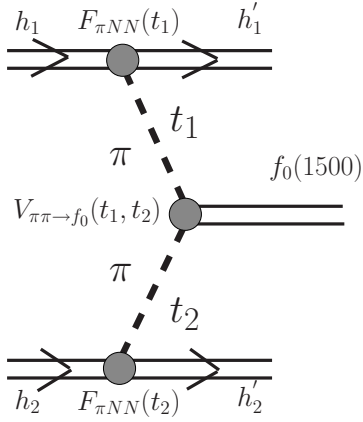}}
   \caption{\label{fig:pion_pion}
   \small  The sketch of the pion-pion MEC mechanism.
Form factors appearing in different vertices and kinematical variables
are shown explicitly.}
\end{figure}

At lower energies ($\sqrt{s} <$ 20 GeV) other processes may 
become important as well.
Since the two-pion channel is one of the dominant decay channels
of $f_0(1500)$ (34.9 $\pm$ 2.3 \%) \cite{PDG}
one may expect the two-pion fusion (see Fig.\ref{fig:pion_pion})
to be one of the dominant mechanisms of exclusive $f_0(1500)$ 
production at the FAIR energies. The two-pion fusion can be
also relative reliably calculated in the framework of 
meson exchange theory. The pion coupling to the nucleon is well
known \cite{Ericson-Weise}. The $\pi NN$ form factor for larger
pion virtualities is somewhat less known. This may limit our
predictions close to the threshold, where rather large 
virtualities are involved due to specific kinematics. 
At largest HESR (antiproton ring) energy, as will be discussed 
in the present paper, this is no longer a limiting factor 
as average pion virtualities are rather small. 

\begin{figure}[!htb]    
\includegraphics[width=0.4\textwidth]{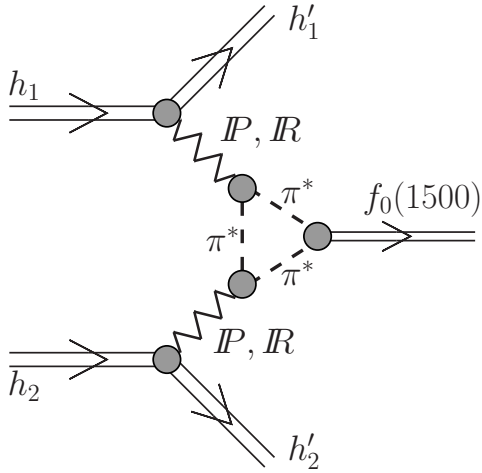}
   \caption{\label{fig:diffractive_pion_triangle}
   \small  The sketch of the double-diffractive mechanism with pionic 
loop for exclusive production of the glueball 
candidate $f_0(1500)$. 
The stars attached to $\pi$ mesons
denote the fact that they are off-mass-shell.}
\end{figure}
In this paper we concentrate on the mechanism of the reaction.
Our aim here is to explore a possibility of studying
exclusive $f_0(1500)$ meson production in the FAIR
and J-PARC energy range and explore the potential
of these facilities. While their are some
ideas about the reaction mechanism at higher energies,
the mechanism at lower energies was never studied.
We shall investigate new mechanisms of pion-pion
fusion shown in Fig.\ref{fig:pion_pion}, 
the QCD mechanism shown in Fig.\ref{fig:QCDdiff} and 
a mechanism with intermediate pionic loop
shown in Fig.\ref{fig:diffractive_pion_triangle}.
The second (QCD) mechanism is typical for high energies
but here we wish to investigate its role at intermediate
energies and in particular its vanishing at low energies
and the interplay with the pion-pion fusion mechanism.

\section{Exclusive processes}

\subsection{Cross section and phase space}

The cross section for a general 3-body reaction $pp\to ppf_0(1500)$
can be written as
\begin{eqnarray}
d\sigma_{pp\to ppM}=\frac{1}{2\sqrt{s (s - 4 m^2)}}\,
\overline {|{\cal M}|^2} \cdot d^{\,3}PS \, . 
\label{general_cross_section_formula}
\end{eqnarray}
Above $m$ is the mass of the nucleon.

The three-body phase space volume element reads
\begin{eqnarray}
d^3 PS = \frac{d^3 p_1'}{2 E_1' (2 \pi)^3} \frac{d^3 p_2'}{2 E_2'
(2 \pi)^3} \frac{d^3 P_M}{2 E_M (2 \pi)^3} \cdot (2 \pi)^4
\delta^4 (p_1 + p_2 - p_1' - p_2' - P_M) \; . 
\label{dPS_element}
\end{eqnarray}
At high energies and small momentum transfers the phase space
volume element can be written as \cite{KMV99}
\begin{eqnarray}
d^3 PS \approx \frac{1}{2^8 \pi^4} dt_1 dt_2 d\xi_1 d\xi_2 d \phi \;
\delta \left( s(1-\xi_1)(1-\xi_2)-M^2 \right) \; ,
\label{dPS_element_he1}
\end{eqnarray}
where $\xi_1$, $\xi_2$ are longitudinal momentum fractions carried
by outgoing protons with respect to their parent protons and the
relative angle between outgoing protons $\phi \in (0, 2\pi)$.
Changing variables  $(\xi_1, \xi_2) \to (x_F, M^2)$ one gets
\begin{eqnarray}
d^3 PS \approx \frac{1}{2^8 \pi^4} dt_1 dt_2 \frac{dx_F}{s \sqrt{x_F^2 +
4 (M^2+|{\bf P}_{M,t}|^2)/s}} \; d \phi \; .
\label{dPS_element_he2}
\end{eqnarray}

The high-energy formulas (\ref{dPS_element_he1}) and
(\ref{dPS_element_he2}) break
close to the meson production threshold.
Then exact phase space formula (\ref{dPS_element}) must be taken 
and another choice of variables is more appropriate.
We choose transverse momenta of the outgoing nucleons ($p'_{1t}, p'_{2t}$),
azimuthal angle between outgoing nucleons ($\phi$) and rapidity of the
meson ($y$) as independent kinematically complete variables.
Then the cross section can be calculated as:
\begin{equation}
d \sigma = \sum_{k} {\cal J}^{-1}(p_{1t},p_{2t},\phi,y)|_{k} 
\frac{\overline{ |{\cal M}(p_{1t},p_{2t},\phi,y)|^2 }  }
{2 \sqrt{s (s-4 m^2)}} \; 
\frac{2 \pi}{(2 \pi)^5}
\frac{1}{2 E'_1} \frac{1}{2 E'_2} \frac{1}{2}
\;  p_{1t} p_{2t} d p_{1t} d p_{2t} d \phi dy \; ,
\end{equation}
where $k$ denotes symbolically discrete solutions of the set
of equations for $p'_{1z}$ and $p'_{2z}$:
\begin{eqnarray}
\left\{ \begin{array}{rcl}
\sqrt{s} - E_M  &=&
\sqrt{m_{1t}^2+p_{1z}^{'2}} + \sqrt{m_{2t}^2+p_{2z}^{'2}} \; ,  \\
-P_{Mz} &=& p'_{1z} + p'_{2z}   \; ,
\end{array} \right.
\label{energy_momentum_conservation}
\end{eqnarray}
where $m_{1t}$ and $m_{2t}$ are transverse masses of outgoing nucleons.
The solutions of Eq.(\ref{energy_momentum_conservation})
depend on the values of integration variables:
$p'_{1z} = p'_{1z}(p'_{1t},p'_{2t},\phi,y)$ and
$p'_{2z} = p'_{2z}(p'_{1t},p'_{2t},\phi,y)$.
The extra Jacobian reads:
\begin{equation}
{\cal J}_k =
\left| \frac{p'_{1z}(k)}{\sqrt{m_{1t}^2+p'_{1z}(k)^2}} -
  \frac{p'_{2z}(k)}{\sqrt{m_{2t}^2+p'_{2z}(k)^2}} \right| \; .
\end{equation}
In the limit of high energies and central production, i.e. $p'_{1z} \gg$ 0 (very forward nucleon1), 
$-p'_{2z} \gg$ 0 (very backward nucleon2) the Jacobian becomes a constant
${\cal J} \to \tfrac{1}{2}$.

The matrix element depends on the process and is a function of
kinematical variables. The mechanism of the exclusive production
of $f_0(1500)$ close to the threshold is not known. 
We shall address this issue here.
Therefore different mechanisms will be considered and the corresponding
cross sections will be calculated.

\subsection{Diffractive QCD amplitude}

According to Khoze-Martin-Ryskin approach (KMR) \cite{KMR},
the amplitude of exclusive double diffractive colour singlet
production $pp\to ppf_0(1500)$ can be written as
\begin{eqnarray}
{\cal
M}^{g^*g^*}=\frac{s}{2}\cdot\pi^2\frac12\frac{\delta_{c_1c_2}}{N_c^2-1}\,
\Im\int
d^2 q_{0,t}V^{c_1c_2}_J\frac{f^{off}_{g,1}(x_1,x_1',q_{0,t}^2,
q_{1,t}^2,t_1)f^{off}_{g,2}(x_2,x_2',q_{0,t}^2,q_{2,t}^2,t_2)}
{q_{0,t}^2\,q_{1,t}^2\, q_{2,t}^2}. \label{ampl}
\end{eqnarray}
The normalization of this amplitude differs from the KMR one
\cite{KMR} by the factor $s/2$ and coincides with the
normalization in our previous work on exclusive $\eta'$-production
\cite{SPT07}. The amplitude is averaged over the colour indices
and over two transverse polarisations of the incoming gluons
\cite{KMR}. The bare amplitude above is subjected to absorption
corrections which depend on collision energy
(the bigger the energy, the bigger the absorption corrections). 
We shall discuss this issue shortly when presenting our results.

The vertex factor $V_J^{c_1c_2}=V_J^{c_1c_2}(q_{1,t}^2,
q_{2,t}^2,P_{M,t}^2)$ in expression (\ref{ampl}) describes the
coupling of two virtual gluons to $f_0(1500)$
meson. Recently the vertex was obtained for off-shell values
of $q_{1,t}$ and $q_{2,t}$ in the case of $\chi_c(0)$ exclusive
production \cite{PST07}.
An almost alternative way to describe the vertex is to express it via
partial decay width $\Gamma(M \to gg)$. 
\footnote{The last value is not so well known. We shall take
$\Gamma(M \to gg) = \Gamma_M^{tot}$. This will give us an upper estimate. As a consequence this will allow us to show
that the gluonic component is negligible for future experiments with the PANDA detector.}
The latter (approximate) method can be used also for 
the $f_0(1500)$ meson production.

In the original Khoze-Martin-Ryskin (KMR) approach \cite{KMR}
the amplitude is written as
\begin{equation}
{\cal M}=N\int\frac{d^2q_{0,t}P[f_0(1500)]}{q_{0,t}^2q_{1,t}^2q_{2,t}^2}
f_g^{KMR}(x_1,x'_1,Q_{1,t}^2,\mu^2;t_1)f_g^{KMR}(x_2,x'_2,Q_{2,t}^2,\mu^2;t_2)
\; ,
\label{KMR_amplitude}
\end{equation}
where only one transverse momentum is taken into account somewhat
arbitrarily as
\begin{eqnarray}
Q_{1,t}^2=\mathrm{min}\{q_{0,t}^2,q_{1,t}^2\} \; , \qquad
Q_{2,t}^2=\mathrm{min}\{q_{0,t}^2,q_{2,t}^2\} \; ,
\label{glue_momenta}
\end{eqnarray}
and the normalization factor $N$ can be written in terms
of the $f_0(1500)\to gg$ decay width (see below).

In the KMR approach the large meson mass approximation
$M\gg |{\bf q}_{1,t}|,\,|{\bf q}_{2,t}|$ is adopted, so
the gluon virtualities are neglected in the vertex factor
\begin{equation}
P[f_0(1500)]\simeq(q_{1,t}q_{2,t})=(q_{0,t}+p'_{1,t})(q_{0,t}-p'_{2,t}).
\label{KMR_vert}
\end{equation}

The KMR UGDFs are written in the factorized form:
\begin{equation}
f_g^{KMR}(x,x',Q_t^2,\mu^2;t)=f_g^{KMR}(x,x',Q_t^2,\mu^2)\exp(b_0t)
\end{equation}
with $b_0=2$ GeV$^{-2}$ \cite{KMR}. In our approach we use
somewhat different parameterization of the $t$-dependent isoscalar form factors.
%

Please note that the KMR and our (general) skewed UGDFs have different
number of arguments. In the KMR approach there is only one effective
gluon transverse momentum (see Eq.(\ref{glue_momenta})) compared to two
independent transverse momenta in general case (see Eq.(\ref{skewed_UGDFs})).

The KMR skewed distributions are given in terms of conventional
integrated densities $g$ and the so-called Sudakov form factor $T$ as
follows:
\begin{equation}
f_g^{KMR} (x,x',Q_t^2, \mu^2) = R_g
\frac{\partial}{\partial \ln Q_t^2}
\left[
\sqrt{T(Q_t^2,\mu^2)} x g(x,Q_t^2)
\right] \; .
\label{KMR_UGDF}
\end{equation}
The square root here was taken using arguments that only survival
probability for hard gluons is relevant.
It is not so-obvious if this approximation is reliable for light
meson production.
The factor $R_g$ in the KMR approach approximately accounts for 
the single $\log Q^2$ skewed effect \cite{KMR}. 
Please note also that in contrast to our approach
the skewed KMR UGDF does not explicitly depend on $x'$
(assuming $x' \ll x \ll 1$). Usually this factor is estimated to be
1.3--1.5. In our evaluations here we take it to be equal 1 to avoid
further uncertainties.
%
%
Following now the KMR notations we write the total
amplitude (\ref{ampl}) (averaged over colour and
polarisation states of incoming gluons) in the limit $M\gg
q_{1,t},\,q_{2,t}$ as
\begin{eqnarray}
{\cal M}=A\,\pi^2\,\frac{s}{2}\,\int d^2
q_{0,t}P[f_0(1500)]\frac{f^{off}_{g,1}(x_1,x_1',q_{0,t}^2,
q_{1,t}^2,t_1)f^{off}_{g,2}(x_2,x_2',q_{0,t}^2,q_{2,t}^2,t_2)}
{q_{0,t}^2\,q_{1,t}^2\, q_{2,t}^2}\,, \label{ampl-KMR}
\end{eqnarray}
where the normalization constant is
%
%

%
\begin{eqnarray}
A^2 = \frac{64\pi\Gamma(f_0(1500) \rightarrow
gg)}{(N_c^2-1)M_{f_{0}}^3} \; .
\label{A-from-Gamma}
\end{eqnarray}
%
%
%
%
%


In addition to the standard KMR approach we could use other
off-diagonal distributions (for details and a discussion see
\cite{SPT07,PST07}).
In the present work we shall use a few sets of unintegrated gluon
distributions which aim at the description of phenomena where
small gluon transverse momenta are involved. Some details
concerning the distributions can be found in Ref.~\cite{LS06}. We
shall follow the notation there.

In the general case we do not know off-diagonal UGDFs very well.
In \cite{SPT07,PST07} we have proposed a prescription how to calculate
the off-diagonal UGDFs:
\begin{eqnarray}\nonumber
f_{g,1}^{off} &=& \sqrt{f_{g}^{(1)}(x_1',q_{0,t}^2,\mu_0^2) \cdot
f_{g}^{(1)}(x_1,q_{1,t}^2,\mu^2)} \cdot F_1(t_1)\,, \\
f_{g,2}^{off} &=& \sqrt{f_{g}^{(2)}(x_2',q_{0,t}^2,\mu_0^2) \cdot
f_{g}^{(2)}(x_2,q_{2,t}^2,\mu^2)} \cdot F_1(t_2)\,,
\label{skewed_UGDFs}
\end{eqnarray}
where $F_1(t_1)$ and $F_1(t_2)$ are isoscalar nucleon form factors.
They can be parameterized as (\cite{PST07})
\begin{eqnarray}
F_1(t_{1,2}) = \frac{4 m_p^2 - 2.79\,t_{1,2}} {(4 m_p^2
-t_{1,2})(1-t_{1,2}/0.71)^2} \;.
\label{off-diag-formfactors}
\end{eqnarray}
In the following for brevity we shall use notation $t_{1,2}$ 
which means $t_1$ or $t_2$.
Above $t_1$ and $t_2$ are total four-momentum transfers in the first
and second proton line, respectively.
While in the emission line the choice of the scale is rather
natural, there is no so-clear situation for the second screening-gluon
exchange
\cite{SPT07}.

Even at intermediate energies ($W$ = 10-50 GeV) typical 
$x_1^{'} = x_2^{'}$ are relatively small ($\sim$ 0.01).
However, characteristic $x_1, x_2 \sim M_{f_0}/\sqrt{s}$ are not 
too small (typically $>$ 10$^{-1}$). Therefore here we cannot use 
the small-$x$ models of UGDFs. In the latter case a Gaussian smearing
of the collinear distribution seems a reasonable solution:
\begin{eqnarray}
f_{g}^{Gauss}(x,k_t^2,\mu_F^2)=xg^{coll}(x,\mu_F^2) \cdot
F_{Gauss}(k_t^2;\sigma_0)\;,
\label{Gaussian_UGDF}
\end{eqnarray}
where $g^{coll}(x,\mu_F^2)$ are standard collinear (integrated)
gluon distribution and $f_{Gauss}(k_t^2;\sigma_0)$ is a Gaussian
two-dimensional function
\begin{equation}
\begin{split}
F_{Gauss}(k_t^2,\sigma_0)=\frac{1}{2\pi\sigma_0^2} \exp\left(-k_t^2/2
\sigma_0^2\right)/\pi  \,. 
\label{Gaussian}
\end{split}
\end{equation}
Above $\sigma_0$ is a free parameter which one can expect to be of 
the order of 1 GeV. Based on our experience in \cite{SPT07} we expect
strong sensitivity to the actual value of the parameter $\sigma_0$.
Summarizing, a following prescription for the off-diagonal UGDF
seems reasonable:
\begin{equation}
f(x,x',k_t^2,k_t^{'2},t) = \sqrt{
f_{small-x}(x',k_t^{'2}) f_{g}^{Gauss}(x,k_t^2,\mu^2)
} \cdot F(t) \; ,
\label{mixed_prescription}
\end{equation}
where $f_{small-x}(x',k_t^{'2})$ is one of the typical small-$x$ 
UGDFs (see e.g.\cite{LS06}). So exemplary combinations are: 
KL $\otimes$ Gauss, BFKL $\otimes$ Gauss, GBW $\otimes$ Gauss 
(for notation see \cite{LS06}).
The natural choice of the scale is $\mu^2 = M_{f_0}^2$. 
This relatively low scale is possible with the
GRV-type of PDF parameterization \cite{GRV}.
We shall call (\ref{mixed_prescription}) a "mixed prescription"
for brevity.

\subsection{Two-gluon impact factor approach for subasymptotic energies}

The amplitude in the previous section, written in terms of 
off-diagonal UGDFs, was constructed for rather large energies.
The smaller the energy the shorter the QCD ladder. 
It is not obvious how to extrapolate the diffractive amplitude 
down to lower (close-to-threshold) energies.
Here we present slightly different method which seems more adequate
at lower energies.

At not too large energies the amplitude of elastic scattering
can be written as amplitude for two-gluon exchange 
\cite{pp_elastic_2g_if,SNS02}
\begin{equation}
{\cal M}_{pp \to pp}(s,t) =
i s \frac{N_c^2-1}{N_c^2}
\int d^2 k_t \; \alpha_s(k_{1t}^2) \alpha_s(k_{2t}^2)
\frac{ 3 F({\bf k}_{1t},{\bf k}_{2t})
       3 F({\bf k}_{1t},{\bf k}_{2t}) }
{(k_{1t}^2 + \mu_g^2) (k_{2t}^2 + \mu_g^2)} \; .
\label{2g_elastic_amplitude}
\end{equation}
In analogy to dipole-dipole or pion-pion scattering 
(see e.g. \cite{SNS02}) the impact factor can be parameterized as:
\begin{equation}
F({\bf k}_{1t},{\bf k}_{2t}) =
\frac{A^2}{A^2+({\bf k}_{1t} + {\bf k}_{2t})^2} -
\frac{A^2}{A^2+({\bf k}_{1t} - {\bf k}_{2t})^2} \; .
\label{impact_factor}
\end{equation}
At high energy the net four-momentum transfer:
$t = -({\bf k}_{1t}+{\bf k}_{2t})^2$.
$A$ in Eq.(\ref{impact_factor}) is a free parameter which
can be adjusted to elastic scattering. For our rough estimate
we take $A = m_{\rho}$.

\begin{figure}[!htb]    
\includegraphics[width=0.355\textwidth]{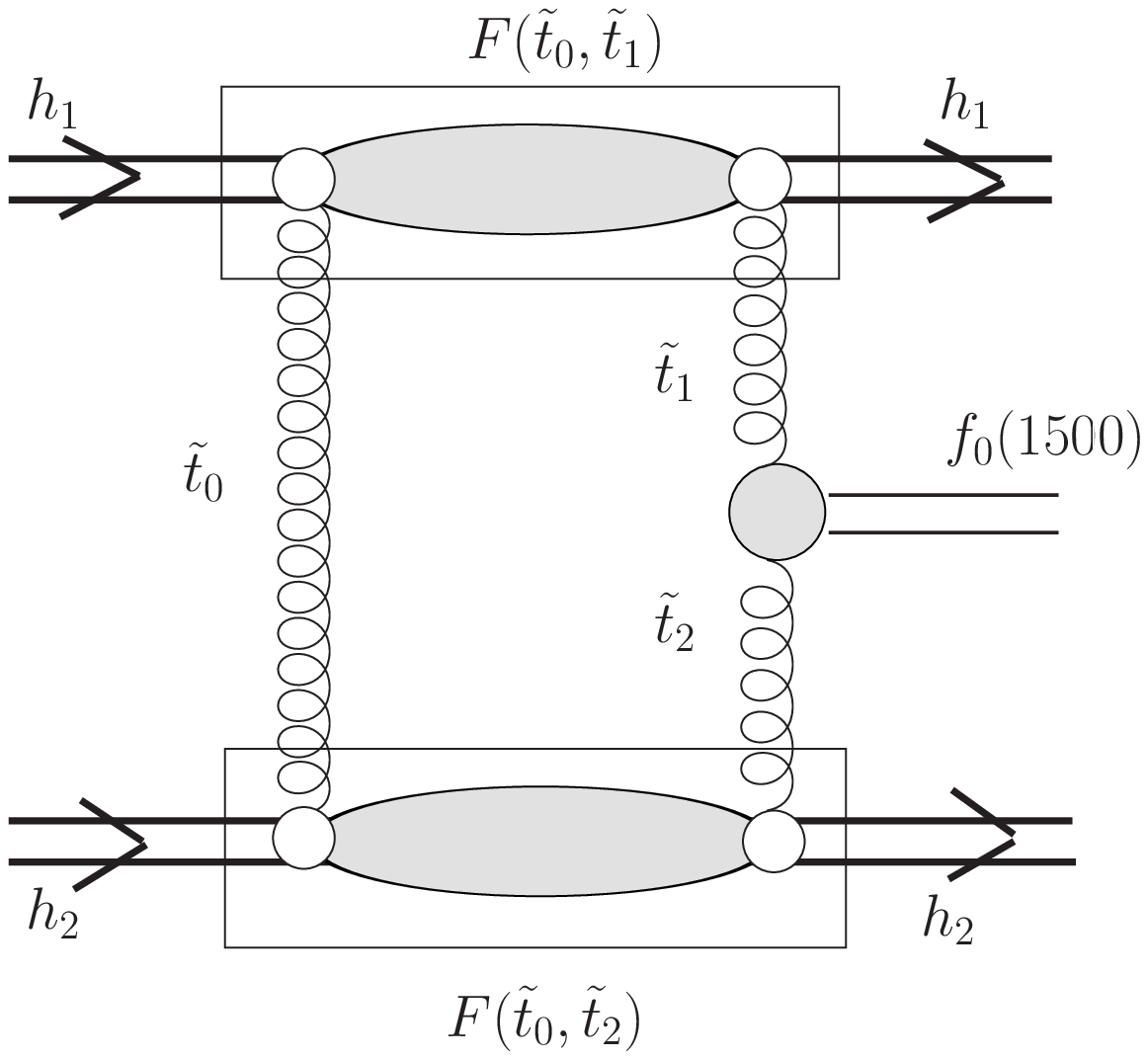}
\includegraphics[width=0.34\textwidth]{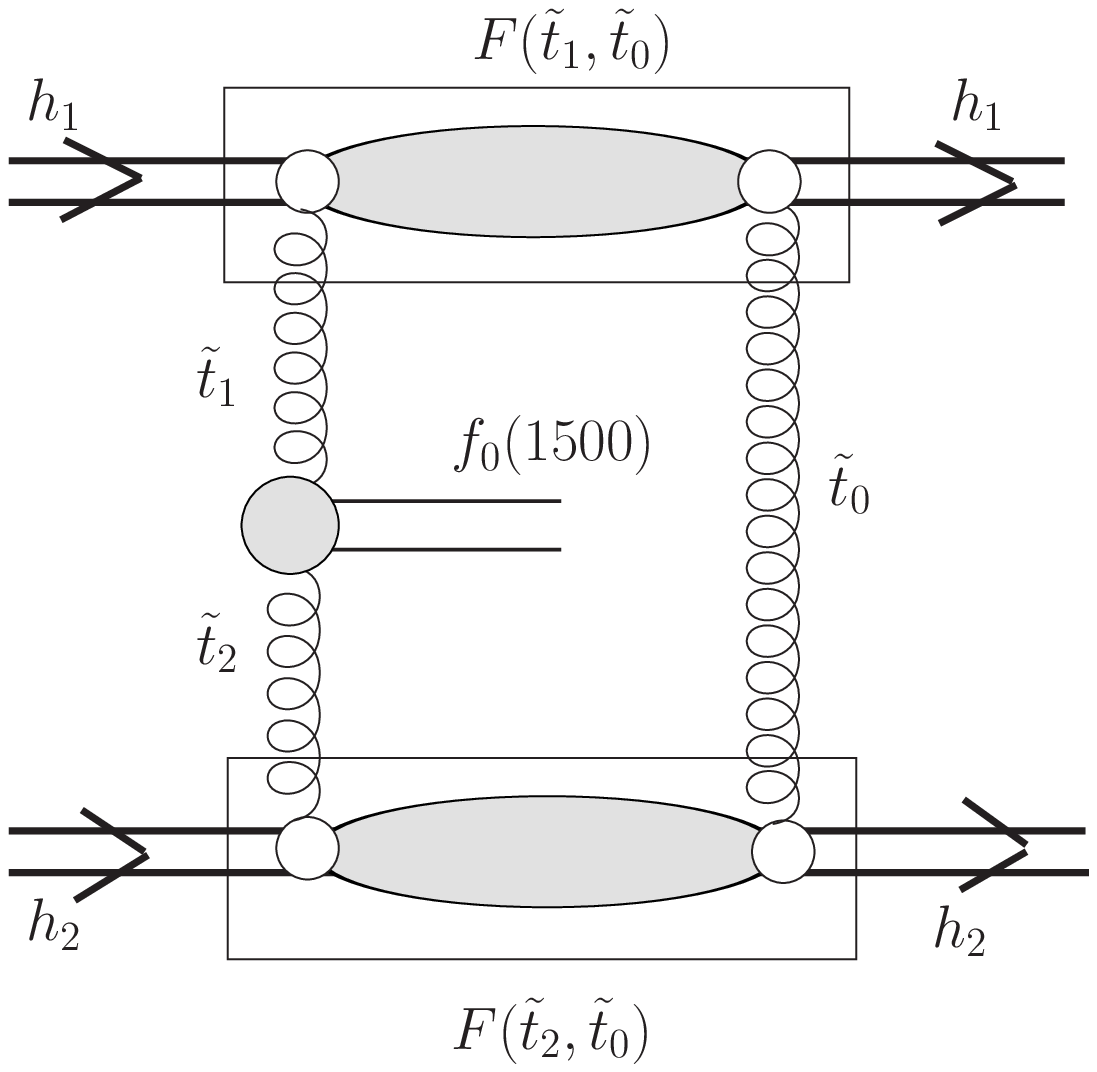}
   \caption{\label{fig:2g_impact_factor}
   \small  The sketch of the two-gluon impact factor approach.
Some kinematical variables are shown explicitly.}
\end{figure}

Generalizing, the amplitude for exclusive $f_0(1500)$ production
can be written as the amplitude for three-gluon exchange shown 
in Fig.\ref{fig:2g_impact_factor}:
\begin{eqnarray}
{\cal M}_{pp \to ppf_0(1500)}(s,y,t_1,t_2,\phi) &=
i s \dfrac{N_c^2-1}{N_c^2} \int  d^2 k_{0t} \;
\left( \alpha_s(k_{0t}^2) \alpha_s(k_{1t}^2) \right)^{1/2}
\left( \alpha_s(k_{0t}^2) \alpha_s(k_{2t}^2) \right)^{1/2} \nonumber \\
 & \times \dfrac{ 3  F({\bf k}_{0t},{\bf k}_{1t}) 3  F({\bf k}_{0t},{\bf k}_{2t}) }
{(k_{0t}^2 + \mu_g^2)
 (k_{1t}^2 + \mu_g^2)
 (k_{2t}^2 + \mu_g^2)} \;
V_{gg \to f_0(1500)}({\bf k}_{1t},{\bf k}_{2t})\;.
\label{3g_exclusive_amplitude}
\end{eqnarray}
At high energy and $y \approx$ 0 the four-momentum transfers can 
be calculated as:\\
$t_1 = -({\bf k}_{0t}+{\bf k}_{1t})^2$,
$t_2 = -({\bf k}_{0t}-{\bf k}_{2t})^2$. \\
At low energy and/or $y \ne$ 0 the kinematics is slightly more 
complicated.
Let us define effective four-vector transfers: 
\begin{eqnarray}
q_1 &=& (p'_1-p_1) = (q_{10},q_{1x},q_{1y},q_{1z}) \; ,  \nonumber \\
q_2 &=& (p'_2-p_2) = (q_{20},q_{2x},q_{2y},q_{2z}) \; .
\label{four-vector_tranfers}
\end{eqnarray}
Then $t_1 \equiv q_1^2 = q_{1l}^2 + q_{1t}^2$ and 
$t_2 \equiv q_2^2 = q_{2l}^2 + q_{2t}^2$.
Close to threshold the longitudinal components 
$q_{1l}^2 = q_{10}^2 - q_{1z}^2 \ll$ 0 and
$q_{2l}^2 = q_{20}^2 - q_{2z}^2 \ll$ 0.
Then the amplitude (\ref{3g_exclusive_amplitude}) must be corrected.
Then also four-vectors of exchanged gluons ($k_0$, $k_1$ and $k_2$)
cannot be purely transverse and longitudinal components must be included
as well.
To estimate the effect we use formula (\ref{3g_exclusive_amplitude})
\footnote{It would be more appropriate to calculate in this case
a four-dimensional integral instead of the two-dimensional one.}
but modify the transferred four momenta of gluons entering
the $g^* g^* \to f_0(1500)$ production vertex:
\begin{eqnarray}
k_1 &=& (0,{\bf k}_{1t},0) \to (q_{10},{\bf k}_{1t},q_{1z}) \; , \nonumber \\
k_2 &=& (0,{\bf k}_{2t},0) \to (q_{20},{\bf k}_{2t},q_{2z}) \; 
\label{4momenta_modification}
\end{eqnarray}
and leave $k_0$ purely transverse. This procedure is a bit arbitrary
but comparing results obtained with formula
(\ref{3g_exclusive_amplitude}) with that from the formula with modified
four-momenta would allow to estimate related uncertainties.

We write the vertex function
$g g \to f_0(1500)$ in the following tensorial form
\footnote{In general, another tensorial forms are also possible.
This may depend on the structure of the considered meson.
In principle, the details depend on the form of 
the vertex. To avoid uncertainties in the $k_t$-factorization
approach we work in the on-shell approximation. 
In the on-shell approximation
(or infinitely heavy meson approximation) the vertex
is expressed through decay width and all vertices
should be equivalent. Even if the off-shell effects
are included we do not expect very different energy 
dependence of the cross section for different 
tensorial forms as due to kinematics only small 
virtualities of gluons enter into game. 
}: 
\begin{equation}
V(k_1,k_2,p_M) = C_{f_0(1500) \to gg} \; g_{\mu \nu} k_1^{\mu} k_2^{\nu} \; .
\end{equation}
The normalization factor is obtained from the decay of $f_0(1500)$
into two soft gluons:
\begin{equation}
|C_{f_0(1500) \to gg}|^2 = \frac{64 \pi}{M_{f_0}^3(N_c^2-1)} 
\Gamma_{f_0(1500) \to g g} \; .
\label{vertex_normalization}
\end{equation}
Of course the partial decay width is limited from above:
\begin{equation}
\Gamma_{f_0(1500) \to gg} < \Gamma_{tot} \; .
\end{equation}

The amplitudes discussed here involve transverse momenta in the
infra-red region. Then a prescription how to extend the perturbative
$\alpha_s(k_t^2)$ dependence to a nonperturbative region
of small gluon virtualities is unavoidable. In the following
$\alpha_s(k_t^2)$ is obtained from an analytic freezing proposed by
Shirkov and Solovtsev \cite{SS97}.

\subsection{Double-diffractive mechanism with intermediate pionic triangle}

\begin{figure}[!htb]    
\includegraphics[width=0.4\textwidth]{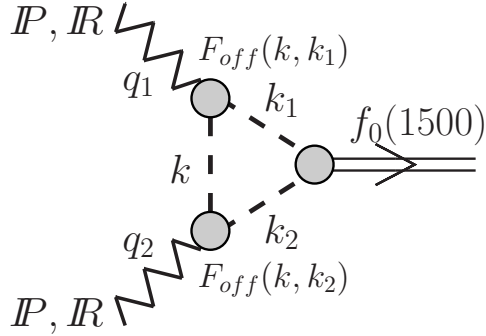}
   \caption{\label{fig:pion_triangle_kinematics}
   \small  A sketch of the double-diffractive mechanism with pionic 
loop for exclusive production of the glueball 
candidate $f_0(1500)$. 
Some kinematical variables are shown explicitly.}
\end{figure}

The $f_0(1500) \to \pi \pi$ is the second most probable
decay channel \cite{PDG}. As a consequence the mechanism
shown in Fig.\ref{fig:pion_triangle_kinematics}
may play important role in the exclusive production
of $f_0(1500)$ \cite{LS09_ddtriangle}. It is relatively
easy to estimate the contribution of this mechanism
at high energies \cite{LS09_ddtriangle}. 
In this paper we shall make an estimate of 
the corresponding cross section not far from the threshold,
where the situation is slightly more complicated.

The amplitude of the process $p p \to p p f_0(1500)$
sketched in Fig.\ref{fig:diffractive_pion_triangle}
can be written in a simplified form \cite{LS09_ddtriangle}
as:
\begin{eqnarray}
&{\cal M}_{\lambda_1 \lambda_2 \to \lambda'_1 \lambda'_2}&( y, p_{1t}, p_{2t}, \phi) \approx  \tilde{T}_{{I\!\!P} {I\!\!P} f_0}(q_1,q_2,p_{f_0})
\delta_{\lambda_1 \lambda'_1} F_{cut}(s_{1,eff}) \delta_{\lambda_2 \lambda'_2} F_{cut}(s_{2,eff})
\nonumber \\
&& \times \left(
   i       s_{1,eff} C_{\pi p}^{{I\!\!P}}\left( \frac{s_{1,eff}}{s_0} \right)^{\alpha_{{I\!\!P}}(t_1)-1}   e^{{\frac{B_{\pi N}}{2}}t_{1}}
+  \eta_f  s_{1,eff} C_{\pi p}^{{I\!\!R}}\left( \frac{s_{1,eff}}{s_0} \right)^{\alpha_{{I\!\!R}}(t_1)-1}   e^{{\frac{B_{\pi N}}{2}}t_{1}}
\right)
\nonumber \\
&& \times \left(
   i       s_{2,eff} C_{\pi p}^{{I\!\!P}}\left( \frac{s_{2,eff}}{s_0} \right)^{\alpha_{{I\!\!P}}(t_2)-1}   e^{{\frac{B_{\pi N}}{2}}t_{2}} \;
+  \eta_f  s_{2,eff} C_{\pi p}^{{I\!\!R}}\left( \frac{s_{2,eff}}{s_0} \right)^{\alpha_{{I\!\!R}}(t_2)-1}   e^{{\frac{B_{\pi N}}{2}}t_{2}} \right).
\nonumber \\
\; 
\label{diffractive_pion_triangle_amplitude}
\end{eqnarray}
The delta functions are related to helicity conservation
in hadronic processes.
While the pomeron (sub)amplitudes are dominantly imaginary,
the reggeon (sub)amplitudes have both real and imaginary parts.
The factor $\eta_{f} \approx i - 1$.
In the formula above
$\alpha_{{I\!\!P}}(t_{1,2}) = \alpha_{{I\!\!P}}(0)
+\alpha_{{I\!\!P}}' \cdot  t_{1,2}$ and 
$\alpha_{{I\!\!R}}(t_{1,2}) = \alpha_{{I\!\!R}}(0)
+\alpha_{{I\!\!R}}' \cdot  t_{1,2}$
are a so-called reggeon trajectories.
For brevity we use notation $t_{1,2}$ 
which means $t_1$ or $t_2$.
We take from the phenomenology:
$\alpha_{{I\!\!P}}(0)$ = 1.0808, 
$\alpha_{{I\!\!P}}'$ = 0.25 GeV$^{-2}$,
$\alpha_{{I\!\!R}}(0)$ = 0.5475 and 
$\alpha_{{I\!\!R}}'$ = 0.93  GeV$^{-2}$ \cite{DL92}.
The strength parameters for the $\pi N$ scattering
fitted to the corresponding total cross sections 
\cite{DL92}:
$C_{\pi p}^{{I\!\!P}}$ = 13.63 mb and $C_{\pi p}^{{I\!\!R}} = (27.56 + 36.02)/2$ mb.
\footnote{We take average value for $\pi^{+}p$ and $\pi^{-}p$ scattering.}
At not too high energies the slope parameter $B_{\pi N}\approx$ 6 GeV$^{-2}$.
The subchannel Mandelstam variable $s_{1,eff}$ and 
$s_{2,eff}$
are related to center-of-mass energies of relevant 
pion-nucleon subsystem. In principle, they are functions
of pion-four momenta in the triangle: 
$s_{1,eff} = s_{1,eff}(k,k_1,p'_1)$
and $s_{2,eff} = s_{2,eff}(k,k_2,p'_2)$
and in general should be put inside of the triangle 
function $T_{{I\!\!P} {I\!\!P} f_0}(q_1,q_2,p_{f_0})$ which depend on 
the 3-body kinematics, i.e. on four-momenta of 
the two exchanged pomerons (in general pomeron-reggeon, reggeon-pomeron or reggeon-reggeon).
In order to simplify the calculation we take instead average values calculated as:
\begin{eqnarray}
s_{1,eff} = (p'_1 + p_{f_0}/2)^2 \; , \nonumber \\
s_{2,eff} = (p'_2 + p_{f_0}/2)^2 \; .
\label{s1_s2}
\end{eqnarray}
The factors $F_{cut}({\hat s})$ are added to cut off
the region of small ${\hat s}$, where the naive Regge 
parametrization does not apply.
We parametrize them in terms of the smooth function:
\begin{eqnarray} 
F_{cut}(\hat s)=\frac{\exp \left( \frac{\hat W-W_{cut}}{a_{cut}}\right)}{1+\exp \left( \frac{\hat W-W_{cut}}{a_{cut}}\right)} \;,
\label{WpiN_correction_factor}
\end{eqnarray}
where ${\hat W} = \sqrt{{\hat s}}$.
The parameter $W_{cut}$ gives the position of the cut and 
parameter $a_{cut}$ describes how sharp is the cut off. 
The latter parameter can have significant influence
on the numerics. We shall take $W_{cut}=2$ GeV
and $a_{cut}=0.1$ GeV. For large energies
$F_{cut}(\hat s)\approx$ 1 and close to
kinematical threshold $\hat W=m_{\pi}+M_{N}:$
$F_{cut}(\hat s)\approx$ 0.
This means that we limit to double-diffractive
contribution only.

The effective Regge parametrizations of $\pi N$ interactions
\cite{DL92} are for both colliding particles
being on-mass-shell.
In our case the triangle pions are off-mass-shell.
We correct the Regge strength parameters by multiplying by two vertex form factors $F_{off}(k,k_{i})$
(see Fig.\ref{fig:pion_triangle_kinematics}).
We take them in the following factorized form:
\begin{eqnarray}
F_{off}(k,k_{i}) = \exp \left( - |k^2-m_{\pi}^2|     / \Lambda_{off}^2 \right) \cdot
                   \exp \left( - |k_{i}^2-m_{\pi}^2| / \Lambda_{off}^2 \right) \; .
\label{triangle_off_formfactors}
\end{eqnarray}
$\Lambda_{off}$ is in principle a free parameter. In the calculation shown
in the result section we shall take $\Lambda_{off}$ = 1 GeV.
The dependence on triangle four-momenta forces us to merge the form factors inside
the triangle integration which leads to a modified
pion-triangle function:
\begin{eqnarray}
\tilde{T}_{{I\!\!P} {I\!\!P} f_0}(q_1, q_2, p_{f_0}) =
\int \frac{d^4 k}{(2 \pi)^4} \hat{T}(k; q_1, q_2, p_{f_0})F_{off}(k,k_{1})F_{off}(k,k_{2})  \; ,
\label{triangle_function}
\end{eqnarray}
where standard triangle integrand $\hat{T}(k; q_1, q_2, p_{f_0})$ reads:
\begin{eqnarray}
\hat{T}(k; q_1, q_2, p_{f_0}) =
&& F(q_1,k_1,k) \frac{1}{(q_1-k)^2 - m_{\pi}^2 + i \epsilon} 
    \nonumber \\
\times && F(q_2,k_2,k) \frac{1}{(q_2+k)^2 - m_{\pi}^2 + i \epsilon}
    \nonumber \\
\times && F(k_1,k_2,p_{f_0}) \; g_{\pi \pi f_0} \;\frac{1}{k^2 - m_{\pi}^2 + i \epsilon}  \; .
\label{triangle_function_hat}
\end{eqnarray}
In addition to three pion propagators we have
written three vertex form factors which are
functions of four momenta of corresponding legs.
In principle, these functions are relatively well known 
for space-like pions.
We parametrize the triangle-vertex form factors
in the following factorized exponential form:
\begin{eqnarray}
F(q_1,k_1,k) = \exp \left( - |k_1^2-m_{\pi}^2| / \Lambda_\pi^2 \right) \cdot
               \exp \left( - |k^2-m_{\pi}^2| / \Lambda_\pi^2 \right)
\; , \nonumber \\
F(q_2,k_2,k) = \exp \left( - |k_2^2-m_{\pi}^2| / \Lambda_\pi^2 \right) \cdot
               \exp \left( - |k^2-m_{\pi}^2| / \Lambda_\pi^2 \right)
\; , \nonumber \\
F(k_1,k_2,p_{f_0}) = \exp \left( - |k_1^2-m_{\pi}^2| / \Lambda_\pi^2 \right) \cdot
                     \exp \left( - |k_2^2-m_{\pi}^2| / \Lambda_\pi^2 \right)
\; .
\label{triangle_formfactors}
\end{eqnarray}
In this factorized form each exponent is associated
with individual leg in the vertex.
Such form factors (exponents) are normalized to unity 
when pions in the loop are on-mass shell.
Please note that we symmetrically 
(modulus in (\ref{triangle_formfactors})) 
damp configurations above and below pion-mass shell.
$\Lambda_{\pi}$ is related to the size of the pions in the triangle.
It is natural to expect: $\Lambda_{\pi} < \Lambda_{off}$.
In the calculation presented here we shall take 
$\Lambda_\pi$ = 0.5 GeV.
Since the configurations close to the mass shells give 
the biggest contributions the sensitivity to the actual 
value of the form factor $F$ (see Eqs.(\ref{triangle_function_hat}) and (\ref{triangle_formfactors})) is not substantial.
The $g_{f_0 \pi \pi}=g_{\pi \pi f_0}$ coupling constant can be calculated
from the corresponding partial decay width \cite{LS09_ddtriangle}.

Calculating the triangle function for running kinematics
of the $h_1 h_2 \to h'_1 h'_2 f_0(1500)$ process
(each point of the phase space) is in practice impossible.
We calculate numerically the triangle function for:
\begin{eqnarray}
q_1 &\to& \left( \left\langle q_{10}\right\rangle_{y=0},0,0,\left\langle q_{1z}\right\rangle_{y=0}\right)  \; , \nonumber \\
q_2 &\to& \left( \left\langle q_{20}\right\rangle_{y=0},0,0,\left\langle q_{2z}\right\rangle_{y=0}\right) \; .
\label{replacements}
\end{eqnarray}
Transverse components are on average small and
are neglected in the present approximation.
Close to threshold $|\left\langle q_{10}\right\rangle_{y=0}|\neq|\left\langle q_{1z}\right\rangle_{y=0}|$
and $|\left\langle q_{20}\right\rangle_{y=0}|\neq|\left\langle q_{2z}\right\rangle_{y=0}|$.

\subsection{Pion-pion MEC amplitude}

It is straightforward to evaluate the pion-pion meson exchange
current contribution shown in Fig.\ref{fig:pion_pion}.
If we assume the $i \gamma_5$ type coupling of the pion to the nucleon 
then the Born amplitude squared
and averaged over initial and summed over final spin 
polarizations reads:
\begin{eqnarray}
\overline{|{\cal M}|^2}=\dfrac{1}{4}&
\left[ \left( E_1 + m \right)\left( E_1'+ m \right)
\left(\dfrac{{\bf p}_1^2}{(E_1 + m)^2} + \dfrac{{\bf p}_1^{'2}}{(E_1^{'} + m)^2} -
\dfrac{ 2 {\bf p}_1 \cdot {\bf p}^{'}_1}{(E_1 + m)(E_1^{'} + m)}\right)\right] \times 2 \nonumber \\
  \times &
\dfrac{g^2_{{\pi NN}}  T_k} {(t_1 - m_{\pi}^2)^2} F_{\pi NN}^2(t_1)
\; \times \; |C_{f_0(1500) \to \pi\pi}|^2  V_{\pi \pi \to f_0(1500)}^2(t_1,t_2) \; \times \; 
\dfrac{g^2_{{\pi NN}}  T_k} {(t_2 - m_{\pi}^2)^2} F_{\pi NN}^2(t_2) \nonumber \\
  \times &
\left[ \left( E_2 + m \right)\left( E_2' + m \right)
\left(\dfrac{{\bf p}_2^2}{(E_2 + m)^2} + \dfrac{{\bf p}_2^{'2}}{(E_2^{'} + m)^2} -
\dfrac{ 2 {\bf p}_2 \cdot {\bf p}^{'}_2}{(E_2 + m)(E_2^{'} + m)}\right)\right] \times 2 \; . \nonumber \\
\label{pion_pion_amplitude}
\end{eqnarray}
In the formula above $m$ is the mass of the nucleon,
$E_1, E_2$ and $E_1', E_2'$ are energies of initial and outgoing nucleons,
${\bf p}_1, {\bf p}_2$ and ${\bf p}'_1, {\bf p}'_2$ are the corresponding
three-momenta and $m_{\pi}$ is the pion mass.
The factor $g_{\pi NN}$ is the familiar pion-nucleon coupling constant
and is relatively well known \cite{ELT02} ($\frac{g_{\pi NN}^2}{4 \pi}$ = 13.5 -- 14.6).
In our calculations we take $\frac{g_{\pi NN}^2}{4 \pi}$ = 13.5.
The isospin factor $T_k$ equals 1 for the $\pi^0 \pi^0$ fusion
and equals 2 for the $\pi^+ \pi^-$ fusion. Limiting to nucleons in the final state, in the case of proton-proton
collisions only $p p f_0(1500)$ final state channel is possible and therefore the $\pi^0 \pi^0$ fusion is allowed while in the case
of proton-antiproton collisions both $p \bar p f_0(1500)$ and
$n \bar n f_0(1500)$ final state channels are possible, i.e. 
both $\pi^0 \pi^0$ and $\pi^+ \pi^-$ MEC are allowed. 
In the case of central heavy meson production rather
large transferred four-momenta squared $t_1$ and $t_2$ are involved
and one has to include extended nature of the particles involved 
in corresponding vertices. This is incorporated via $F_{\pi NN}(t_1)$
or $F_{\pi NN}(t_2)$ vertex form factors. The influence of the
t-dependence of the form factors will be discussed in the result
section. In the meson exchange approach \cite{MHE87} they 
are parameterized in the monopole form as
\begin{equation}
F_{\pi N N}(t) = \frac{\Lambda^2 - m_{\pi}^2}{\Lambda^2 - t} \; .
\label{F_piNN_formfactor}
\end{equation} 
Typical values of the form factor parameter are $\Lambda$ = 1.2--1.4 GeV \cite{MHE87},
however the Gottfried Sum Rule violation prefers smaller $\Lambda \approx$
0.8 GeV \cite{GSR}.

The normalization constant $|C|^2$ in (\ref{pion_pion_amplitude})
can be calculated from the partial decay width as
\begin{equation}
|C_{f_0(1500) \to \pi \pi}|^2 
= \frac{8 \pi \; 2 M_{f_0}^2 \Gamma_{f_0(1500) \to \pi^0 \pi^0}}
{\sqrt{M_{f_0}^2 - 4 m_{\pi}^2 }}   \; ,
\label{normalization_constant}
\end{equation}
where $\Gamma_{f_0(1500) \to \pi^0 \pi^0} = 
0.109 \cdot  BR(f_0(1500) \to \pi \pi) \cdot 0.5$ GeV. The branching
ratio is $BR(f_0(1500) \to \pi \pi)$ = 0.349 \cite{PDG}.
The off-shellness of pions is also included for the 
$\pi \pi \to f_0(1500)$ transition through the extra 
$V_{\pi \pi \to f_0(1500)}(t_1,t_2)$ form factor 
which we take in the factorized form:
\begin{equation}
V_{\pi \pi \to f_0(1500)}(t_1,t_2) = 
\frac{\Lambda_{\pi\pi f_0}^2 - m_{\pi}^2}{\Lambda_{\pi\pi f_0}^2 - t_1} 
\cdot
\frac{\Lambda_{\pi\pi f_0}^2 - m_{\pi}^2}{\Lambda_{\pi\pi f_0}^2 - t_2}
\; .
\label{pipi2f0_formfactor}
\end{equation}
It is normalized to unity when both pions are on mass shell
\begin{equation}
V(t_1=m_{\pi}^2,t_2=m_{\pi}^2) = 1 \; . 
\end{equation}
In the present calculation we shall take $\Lambda_{\pi\pi f_0}$ = 1.0 GeV.

\section{Results}

\subsection{Gluonic QCD mechanism}

Let us start with the QCD mechanism relevant at higher energies.
We wish to present differential distributions in 
$x_F$, $t_1$ or $t_2$ and relative azimuthal angle $\phi$.
In the following we shall assume: 
$\Gamma_{f_0(1500) \to gg} = \Gamma_{f_0(1500)}^{tot}$ .
This assumption means that our differential distributions
mean upper limit of the cross section. If the fractional
branching ratio is known, our results should be multiplied
by its value.
There are almost no absolutely normalized experimental data
on exclusive $f_0(1500)$ production in the literature, except
of Ref.\cite{kirk}. 
The absolutely normalized data 
of the ABCDHW Collaboration \cite{ABCDHW86} put emphasis 
rather on $f_2(1270)$ production.
In principle, some (model-dependent) information on 
glueball wave function could be obtained from radiative
decays $J/\psi \to \gamma f_0(1500)$
and $\Upsilon \to \gamma f_0(1500)$ \cite{MMP04}. 
The present data are not good enough to provide 
a detailed information on coupling of gluons 
to $f_0(1500)$.

\begin{figure}[!htb]    
\includegraphics[width=0.5\textwidth]{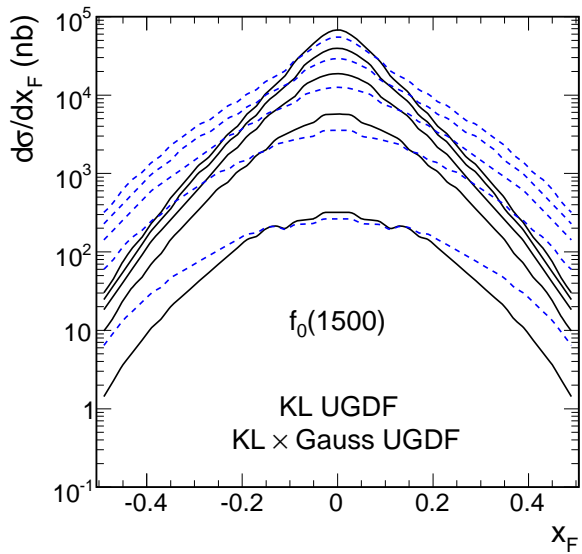}
   \caption{\label{fig:dsig_dxf_kl}
   \small  The distribution of $f_0(1500)$ in Feynman $x_F$ for 
   W = 10, 20, 30, 40, 50 GeV (from bottom to top). In this calculation 
   the Kharzeev-Levin UGDF (solid line) and the mixed distribution 
   KL$\otimes$ Gauss (dashed line) were used.
}
\end{figure}

In Fig.\ref{fig:dsig_dxf_kl} we show as example
distribution in Feynman $x_F$ for Kharzeev-Levin UGDF (solid) and
the mixed distribution KL $\otimes$ Gaussian (dashed)
for several values of collision energy in the interval 
$W$ = 10 -- 50 GeV.
In general, the higher collision energy the larger cross section. 
With the rise of the initial energy the cross section becomes peaked 
more and more at $x_F \sim$ 0. The mixed UGDF produces
slightly broader distribution in $x_F$.


\begin{figure}[!htb]    
\includegraphics[width=0.5\textwidth]{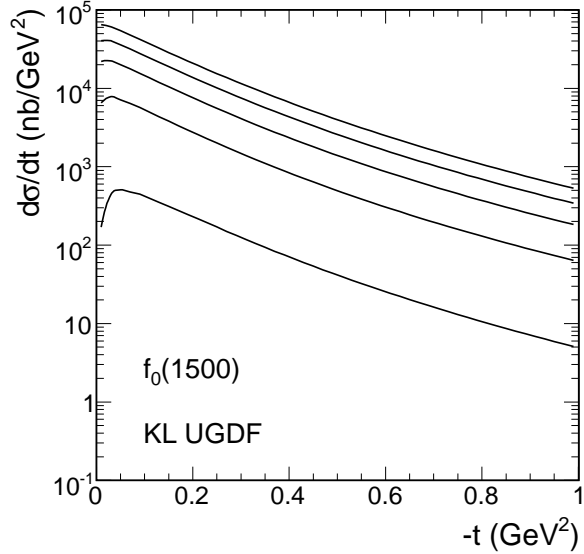}
   \caption{\label{fig:dsig_dt_kl}
   \small  Distribution in $t = t_1 = t_2$ for Kharzeev-Levin UGDF
for $W =$ 10, 20, 30, 40, 50 GeV (from bottom to top).
The notation here is the same as in Fig.\ref{fig:dsig_dxf_kl}.
}
\end{figure}

In Fig.\ref{fig:dsig_dt_kl} we present corresponding distributions
in $t = t_1 = t_2$. The slope depends on UGDF used, but for a given
UGDF is almost energy independent.


Finally we present corresponding distributions in relative 
azimuthal angle between outgoing protons or proton and antiproton 
\footnote{The QCD gluonic mechanism is of course charge independent.}.
These distributions have maximum when outgoing nucleons are back-to-back.
Again the shape seems to be only weekly energy dependent.


\begin{figure}[!htb]    
\includegraphics[width=0.5\textwidth]{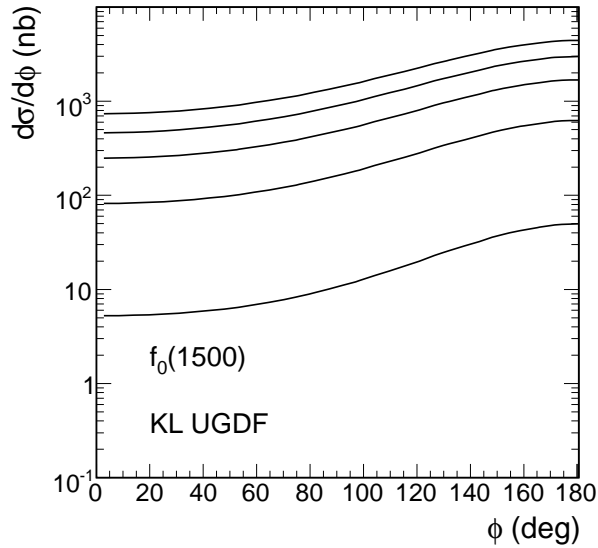}
   \caption{\label{fig:dsig_dphi_kl}
   \small  Distribution in relative azimuthal angle
for Kharzeev-Levin UGDF for $W$ = 10, 20, 30, 40, 50 GeV (from bottom to top).
The notation here is the same as in Fig.\ref{fig:dsig_dxf_kl}.
}
\end{figure}


\subsection{Diffractive versus pion-pion mechanism}

\begin{figure}[!htb]    
\includegraphics[width=0.49\textwidth]{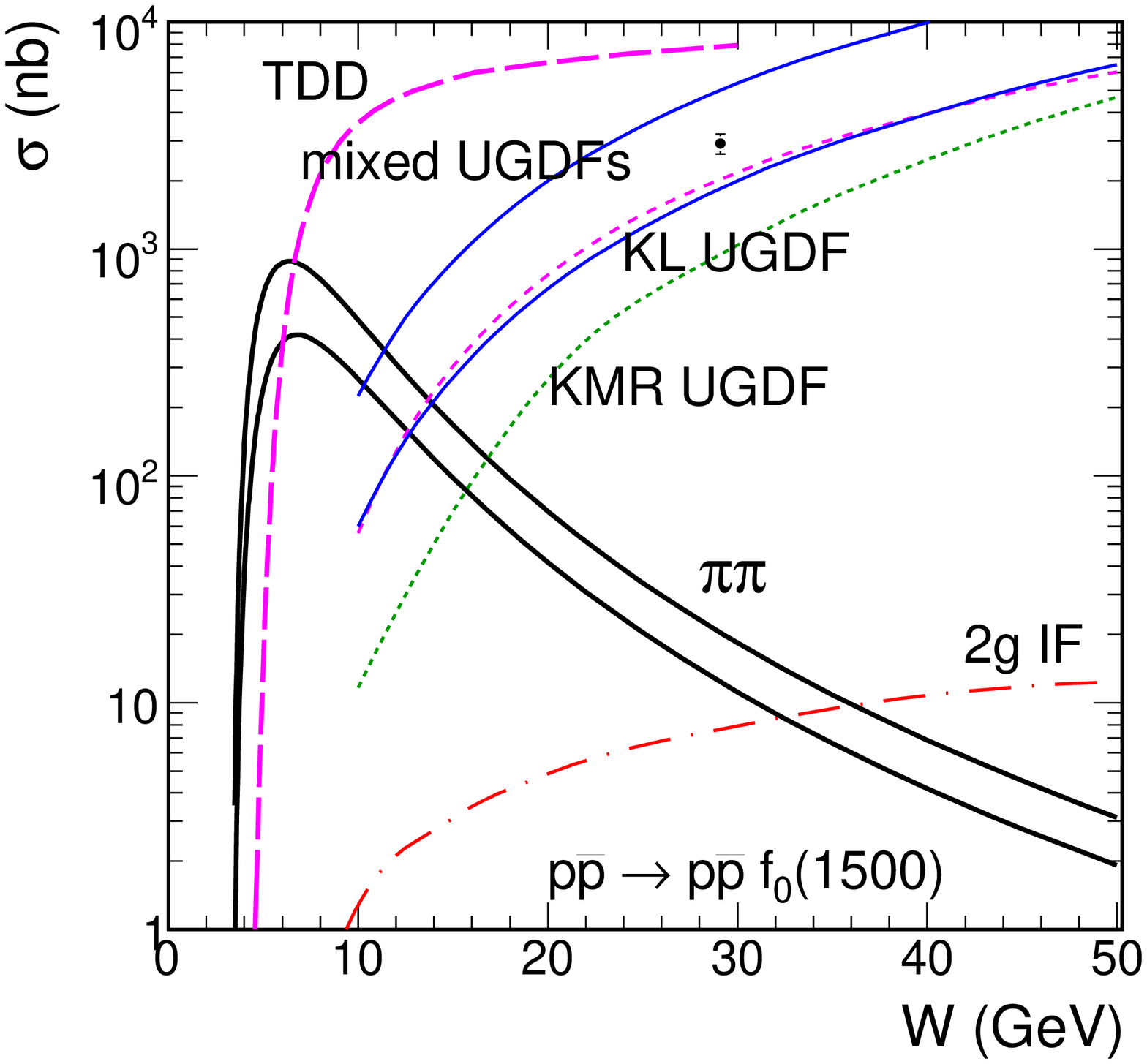}
\includegraphics[width=0.49\textwidth]{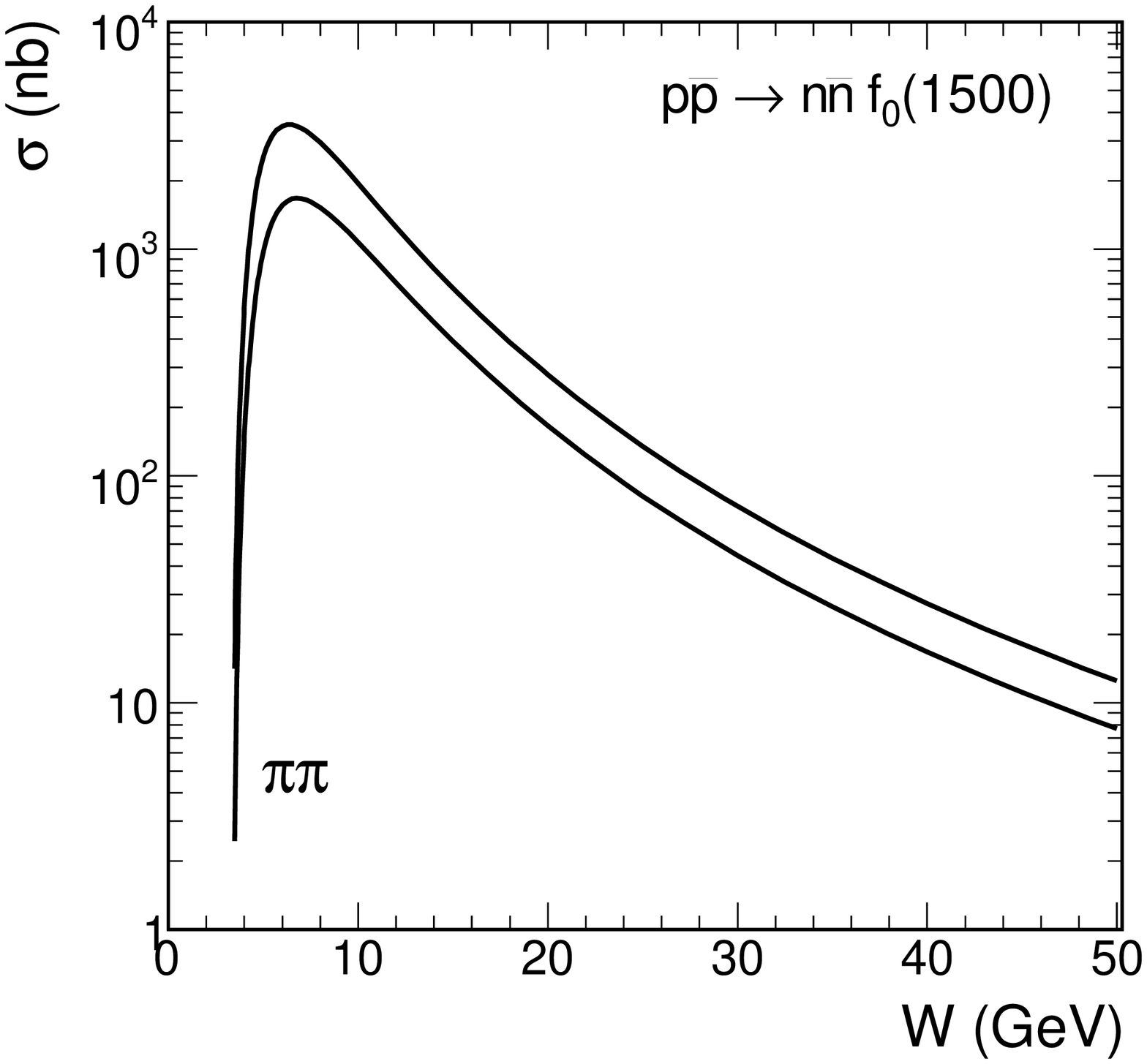}
   \caption{\label{fig:sigma_W}
   \small (Color online.) The integrated cross section as a function
of the center of mass energy for $p \bar p \to p \bar p f_0(1500)$ 
(left panel)
and $p \bar p \to n \bar n f_0(1500)$ (right panel) reactions.
The thick solid lines are for pion-pion MEC contribution ($\Lambda$ = 0.8, 1.2 GeV), the dashed line is 
for QCD diffractive contribution obtained with the Kharzeev-Levin UGDF, 
the dotted line for the KMR approach
and the thin solid lines (blue) are for "mixed" UGDF
(KL $\otimes$ Gaussian) with $\sigma_0$ = 0.5, 1 GeV. 
The dash-dotted line represents the two-gluon impact factor result.
The thick long-dashed line corresponds to the
second diffractive mechanism with intermediate pionic triangle.
The experimental data point at W = 29.1 GeV is from Ref.\cite{kirk}.
}
\end{figure}

What about the pion-pion fusion mechanism? Can it dominate over
the gluonic mechanism discussed in the previous subsection?
In Fig.\ref{fig:sigma_W} we show the integrated cross section
for the exclusive $f_0(1500)$ elastic production
\begin{equation}
p \bar p \to p f_0(1500) \bar p 
\end{equation}
and for double charge exchange reaction
\begin{equation}
p \bar p \to n f_0(1500) \bar n  \; . 
\end{equation}
The thick solid line represents the pion-pion component calculated with 
monopole vertex form factors (\ref{F_piNN_formfactor}) with 
$\Lambda$ = 0.8 GeV (lower) and $\Lambda$ = 1.2 GeV (upper).
The difference between the lower and upper curves represents uncertainties
on the pion-pion component.
The pion-pion contribution grows quickly from the threshold, takes
maximum at $W \approx$ 6-7 GeV and then slowly drops with increasing
energy. The gluonic contribution calculated with unintegrated
gluon distributions drops with decreasing energy 
towards the kinematical threshold and seems to be about order of 
magnitude smaller than the pion-pion component at W = 10 GeV.
We show the result with Kharzeev-Levin UGDF (dashed line) which 
includes gluon saturation effects relevant for small-x, 
Khoze-Martin-Ryskin UGDF (dotted line) used for the exclusive 
production of the Higgs boson and the result with the 
"mixed prescription" (KL $\otimes$ Gaussian) for different
values of the $\sigma_0$ parameter: 0.5 GeV (upper thin solid line),
1.0 GeV (lower thin solid line). In the latter case results rather
strongly depend on the value of the smearing parameter.
The thick dash-dotted line corresponds to the
second diffractive mechanism with pionic triangle.
It is above the WA102 experimental data point.
This is probably because of absorption effects
not included in the present calculation.
This contribution stays below the pion-pion fusion
contribution at the GSI HESR energies.
For comparison we show also experimental data point of the WA102
Collaboration from Ref.\cite{kirk} which lies between the results
obtained with "KL" and "mixed" off-diagonal UGDFs.
The thick long-dashed line corresponds to the
second diffractive mechanism with pionic triangle.
It is above the WA102 experimental data point.
This is probably because of absorption effects
not included in the present calculation.
This contribution stays below the pion-pion fusion
contribution at the GSI HESR energies.

We calculate the gluonic contribution down to W = 10 GeV.
Extrapolating the gluonic component to even lower energies in terms
of UGDFs seems rather unsure. At lower energies the two-gluon impact
factor approach seems more relevant.
The impact factor approach result is even order of magnitude smaller
than that calculated in the KMR approach (see lowest dash-dotted 
(red) line in Fig. \ref{fig:sigma_W}), so it seems that 
the diffractive contribution is completely negligible at 
the FAIR energies.

Our calculation suggests that quite different energy dependence 
of the cross section may be expected in elastic and 
charge-exchange channels. Experimental studies at FAIR and J-PARC
could shed more light on the glueball production mechanism. 

\begin{figure}[!htb]    
\includegraphics[width=0.5\textwidth]{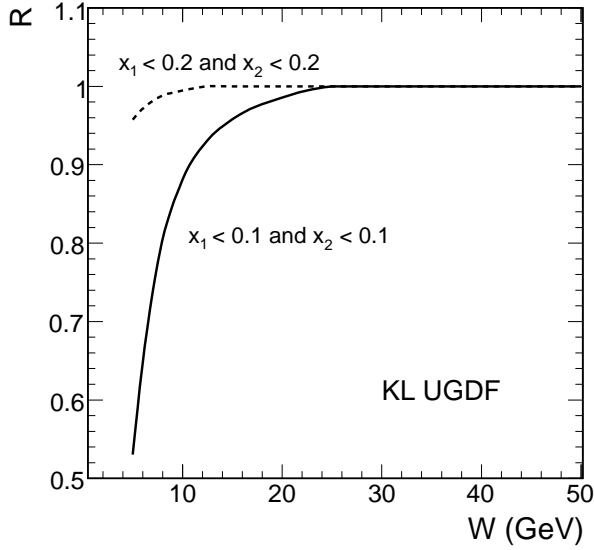}
   \caption{\label{fig:large_x}
   \small  The ratio of the cross sections (see Eq. (\ref{ratio_large_x})) as a function of center-of-mass energy. 
The solid line corresponds to $x_0$ = 0.1 and the dashed line to $x_0$ = 0.2.
}
\end{figure}

The smaller energies the larger values 
of $x_1$ and $x_2$ are involved. 
Many of unintegrated gluon distributions in the literature
are formulated in the region of very small $x$. 
Extrapolation of the method down to small energies 
automatically means going to the region of large $x$. 
Below we wish to demonstrate this fact.
In Fig.\ref{fig:large_x} we show the ratio
of the cross sections
\begin{equation}
R = \frac{\sigma(W;x_1 < x_0, x_2 < x_0)}
                {\sigma(W)} \; ,
\label{ratio_large_x}
\end{equation}
as a function of center-of-mass energy.
Above $x_0$ was introduced to define the region of 
small/large $x$.
The solid line corresponds to $x_0$ = 0.1 and the dashed
line to $x_0$ = 0.2. 
Down to largest HESR energies
one stays in the region of $x_1, x_2 <$ 0.2.

\subsection{Predictions for PANDA at HESR}

Let us concentrate now on $p \bar p$ collisions at 
energies relevant for future experiments at HESR at 
the FAIR facility in GSI.
Here the pion-pion MEC (see Fig.\ref{fig:pion_pion}) seems 
to be the dominant mechanism, especially for the 
charge exchange reaction 
$ p \bar p \to n \bar n f_0(1500) $.
As discussed in the previous section the gluonic component
can be there safely neglected.

\begin{figure}[!htb]    
\includegraphics[width=0.5\textwidth]{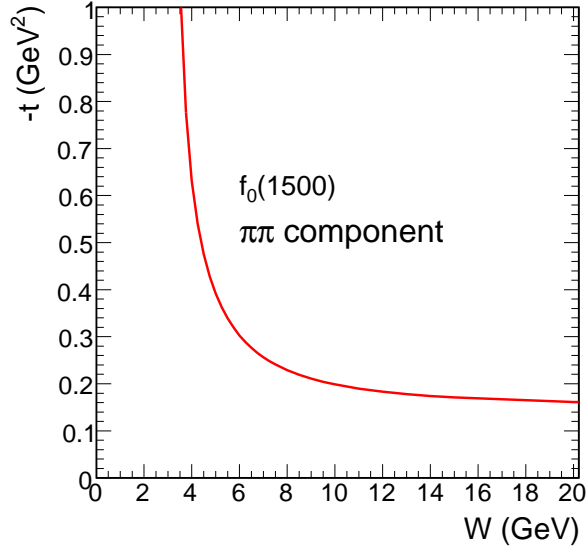}
   \caption{\label{fig:average_t1andt2_W}
   \small (Color online.) 
Average value of $ < t_1 > = < t_2> $ as a function of 
the center-of-mass collision energy for the two-pion exchange mechanism.
In this calculation $\Lambda$ = 0.8 GeV.
}
\end{figure}

In Fig.\ref{fig:average_t1andt2_W} we show average values
of $t_1$ (or $t_2$) for the two-pion MEC as a function of
the center of mass energy.
Close to threshold $W = 2 m_N + m_{f_0(1500)}$ the transferred
four-momenta squared are the biggest, of the order of about 
1.5 GeV$^2$.
The bigger energy the smaller the transferred four-momenta squared.
Therefore experiments close to threshold open a unique 
possibility to study physics of large transferred four-momenta
squared at relatively small energies.
This is a quite new region, which was not studied so far in
the literature.  

Below we shall present cross sections for the
$p \bar p \to n \bar n f_0(1500)$ reaction.
The cross section for the $p \bar p \to p \bar p f_0(1500)$
reaction can be obtained by rescaling by the factor 
of 1/4.


\begin{figure}[!htb]    
\includegraphics[width=0.5\textwidth]{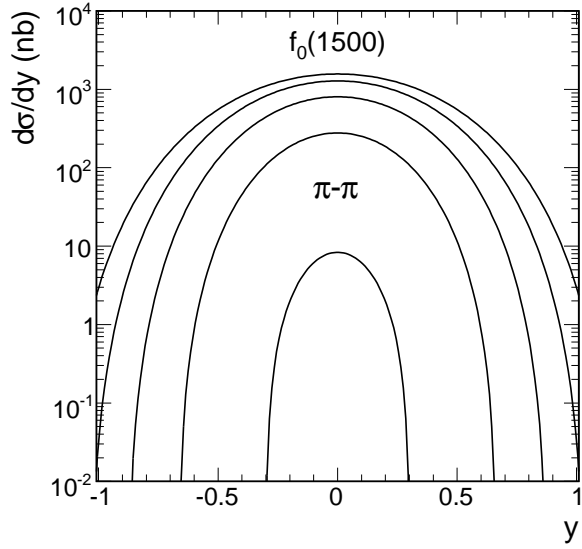}
   \caption{\label{fig:dsig_dy_pipi_f0_1500}
   \small  Rapidity distribution of $f_0(1500)$ ($\pi^+ \pi^-$ fusion only)
produced in the reaction $ p \bar p \to n \bar n f_0(1500) $
for W = 3.5, 4.0, 4.5, 5.0, 5.5 GeV (from bottom to top).
In this calculation $\Lambda$ = 0.8 GeV.
}
\end{figure}

The maximal energy planned for HESR is $\sqrt{s}$ = 5.5 GeV.
At this energy the phase space is still very limited.
In Fig.\ref{fig:dsig_dy_pipi_f0_1500} we show rapidity distribution
of $f_0(1500)$ calculated including pion-pion fusion only. 
For comparison the rapidity of incoming antiproton
and proton is 1.74 and -1.74, respectively. This means that in the
center-of-mass system the glueball is produced at midrapidities,
on average between rapidities of outgoing nucleons.


\begin{figure}[!htb]    
\includegraphics[width=0.5\textwidth]{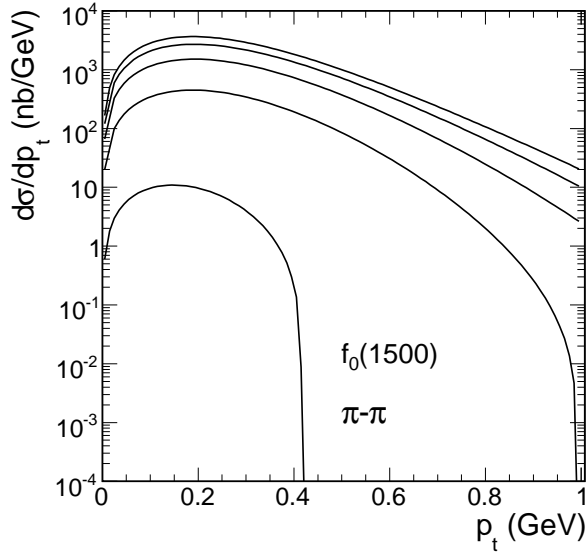}
   \caption{\label{fig:dsig_dpt_pipi_f0_1500}
   \small  Transverse momentum distribution of neutrons or antineutrons
produced in the reaction $ p \bar p \to n \bar n f_0(1500) $ ($\pi^+ \pi^-$
fusion only) 
for W = 3.5, 4.0, 4.5, 5.0, 5.5 GeV (from bottom to top).
In this calculation $\Lambda$ = 0.8 GeV.
}
\end{figure}

In Fig.\ref{fig:dsig_dpt_pipi_f0_1500} we show transverse
momentum distribution of neutrons or antineutrons produced in 
the reaction $ p \bar p \to n \bar n f_0(1500) $. 
The distribution depends on
the $\pi N N$ form factors $F_{\pi NN}(t_1)$ and $F_{\pi NN}(t_2)$ 
in formula (\ref{pion_pion_amplitude}). 


\begin{figure}[!htb]    
\includegraphics[width=0.5\textwidth]{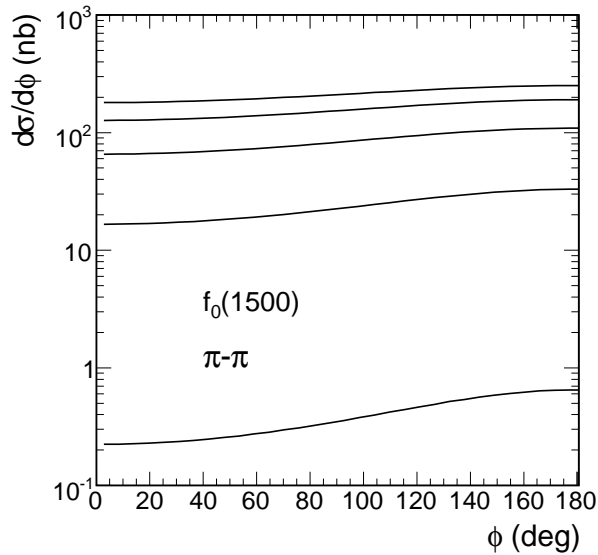}
   \caption{\label{fig:dsig_dphi_pipi_f0_1500}
   \small  Azimuthal angle correlations between neutron and antineutron
produced in the reaction $ p \bar p \to n \bar n f_0(1500) $ 
($\pi^+ \pi^-$ fusion only) 
for W = 3.5, 4.0, 4.5, 5.0, 5.5 GeV (from bottom to top).
In this calculation $\Lambda$ = 0.8 GeV.
}
\end{figure}

In Fig.\ref{fig:dsig_dphi_pipi_f0_1500} we show azimuthal angle
correlation between outgoing hadrons (in this case neutron and
antineutron). The preference for back-to-back configurations is
caused merely by the limitations of the phase space close to 
the threshold (the matrix element for pion-pion fusion is 
$\phi$-independent).
This correlation vanishes in the limit of infinite energy.
At high energy, where the phase space limitations are
small, the distributions are isotropic, there is no
dependence on azimuthal angle.
In practice far from the threshold the distribution becomes almost 
constant in azimuth.  This has to be contrasted with similar 
distributions for pomeron-pomeron fusion shown in 
Fig.\ref{fig:dsig_dphi_kl}
which are clearly peaked for the back-to-back configurations.
Therefore a deviation from the constant distribution in relative 
azimuthal angle for the highest HESR energy of W = 5.5 GeV 
for $p \bar p \to p f_0(1500) \bar p$ can be a signal of the gluon 
induced processes and/or the presence of subleading reggeon exchanges, 
e.g. $\rho \rho$. It is not well understood what happens with 
the gluon induced diffractive processes when going down to 
intermediate (W = 5-10 GeV) energies. Our calculations shows,
however, that the diffractive component is negligible compared
to the pion-pion fusion at $W <$ 10 GeV.
Possible future experiments performed at J-PARC could
bring some new insights into this issue by studying distortions
(probably very small) from the pion-pion fusion mechanism. 

\begin{figure}[!htb]    
\includegraphics[width=0.4\textwidth]{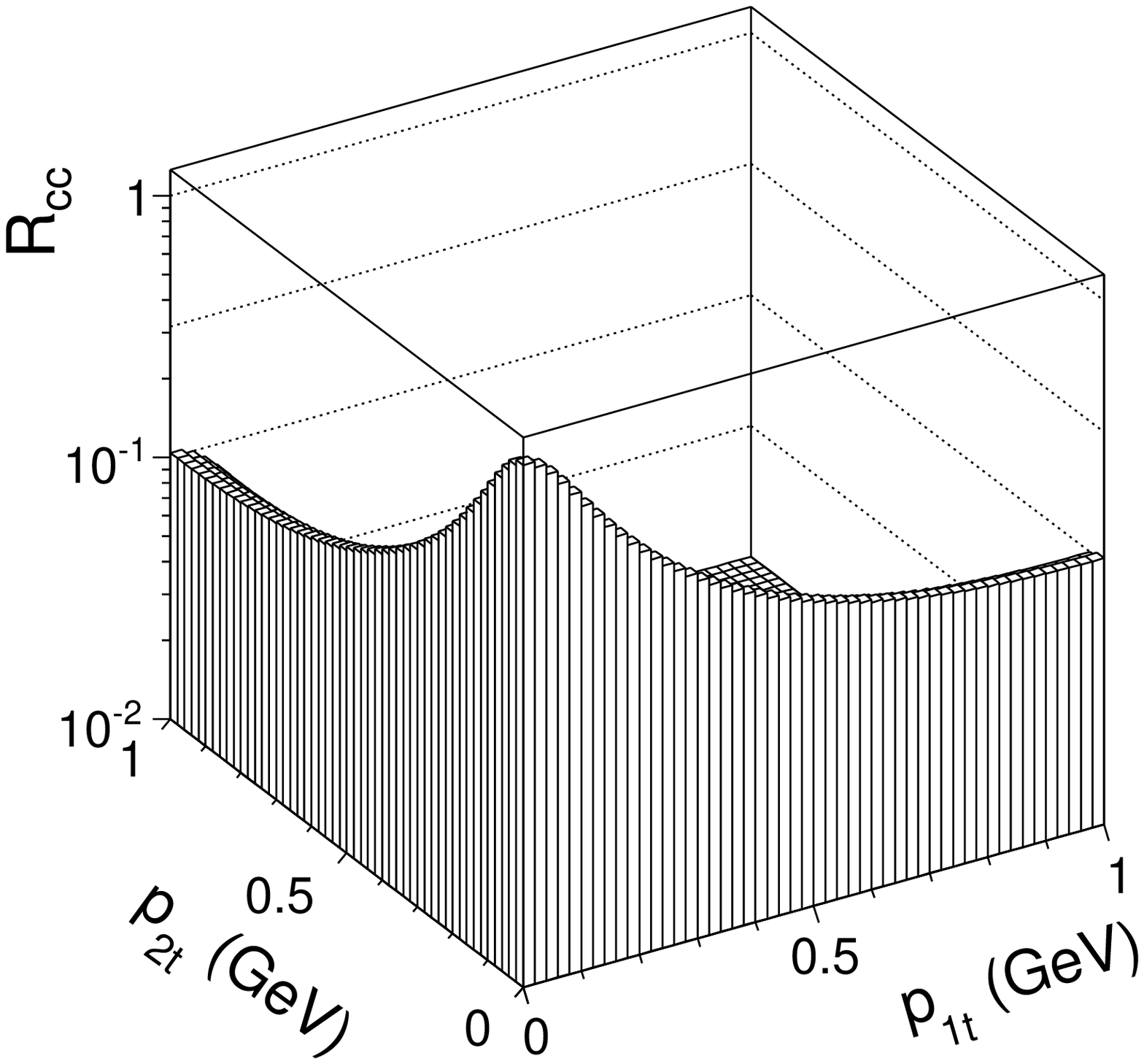}
\includegraphics[width=0.4\textwidth]{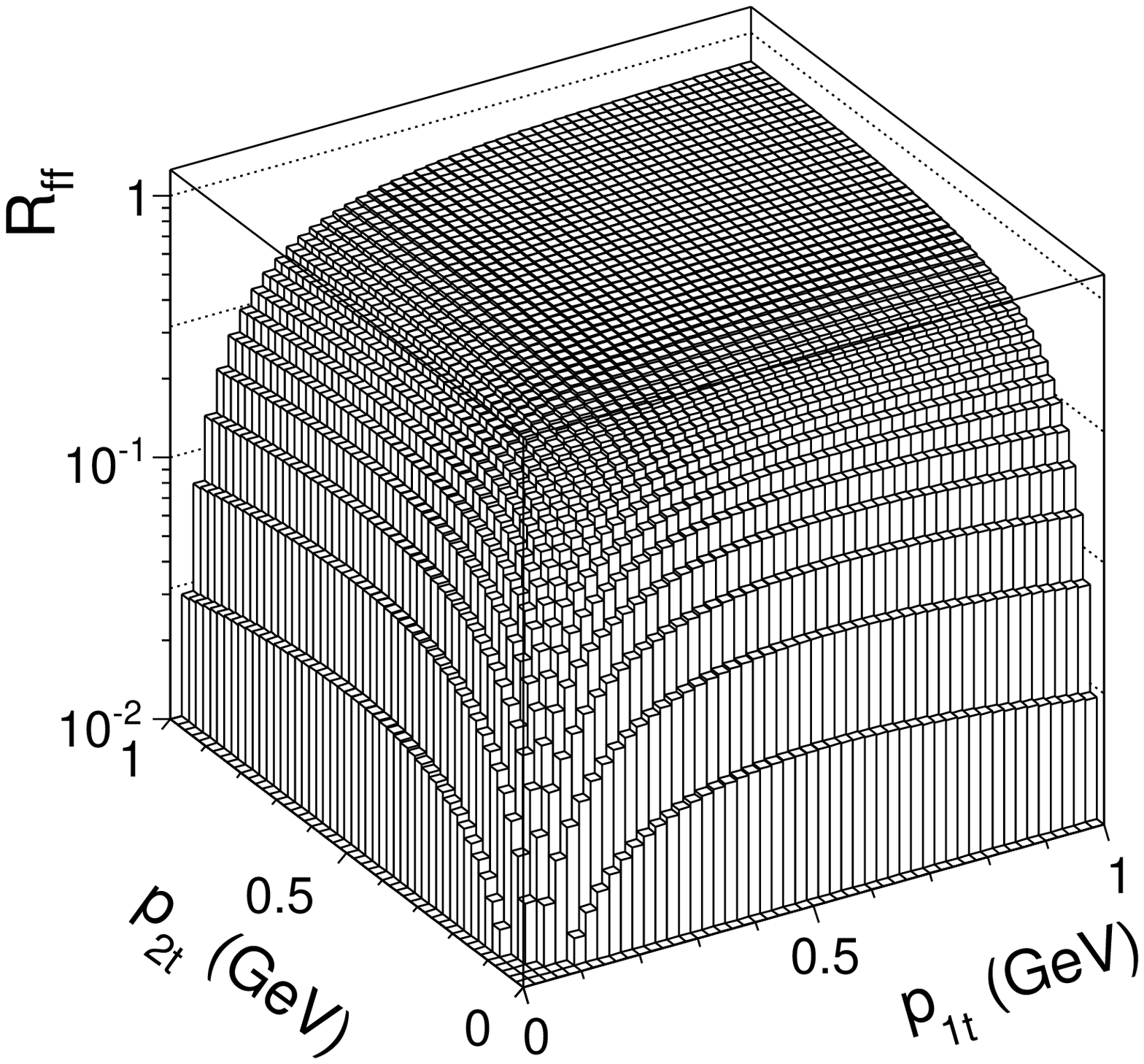}
\includegraphics[width=0.4\textwidth]{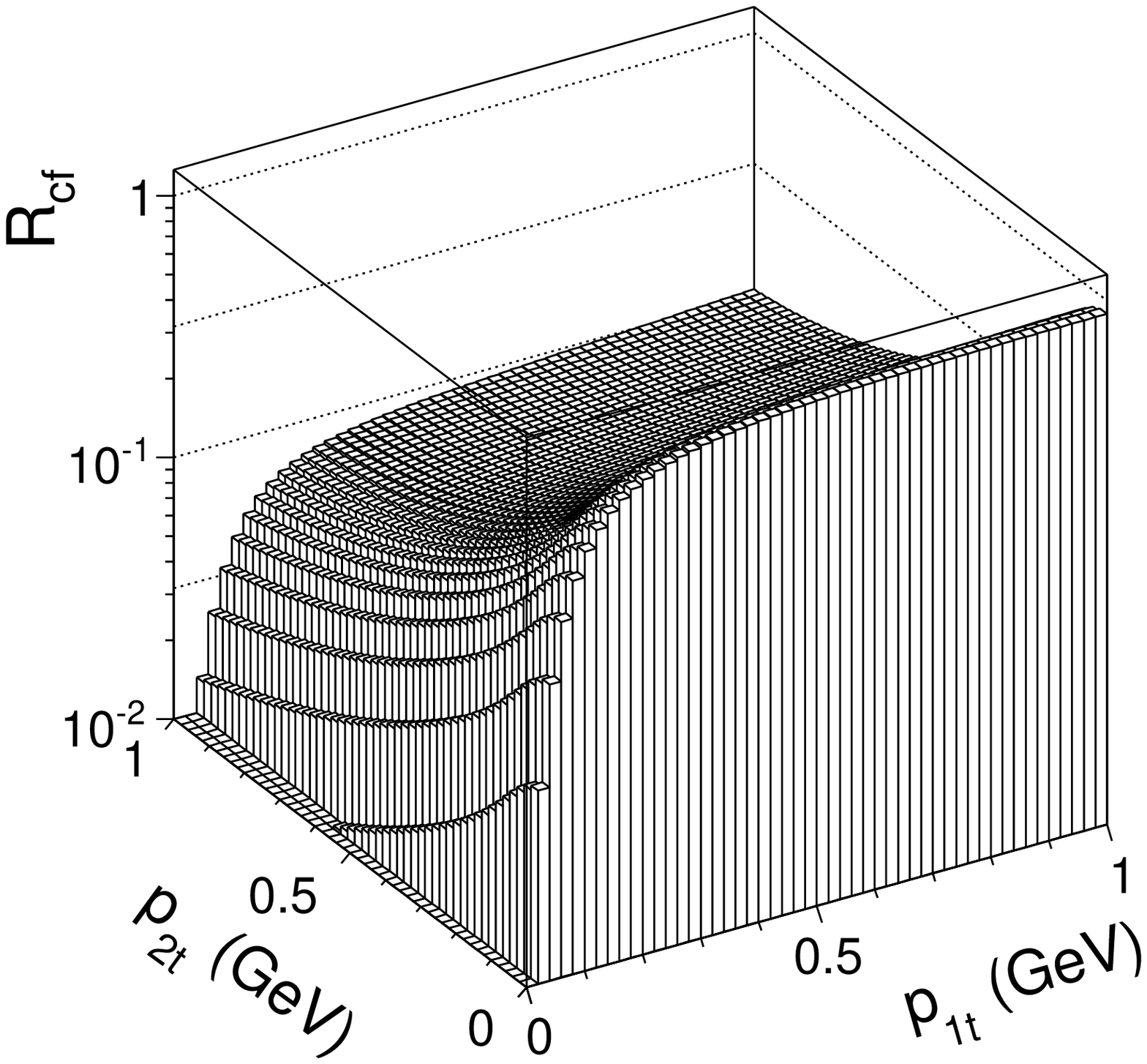}
\includegraphics[width=0.4\textwidth]{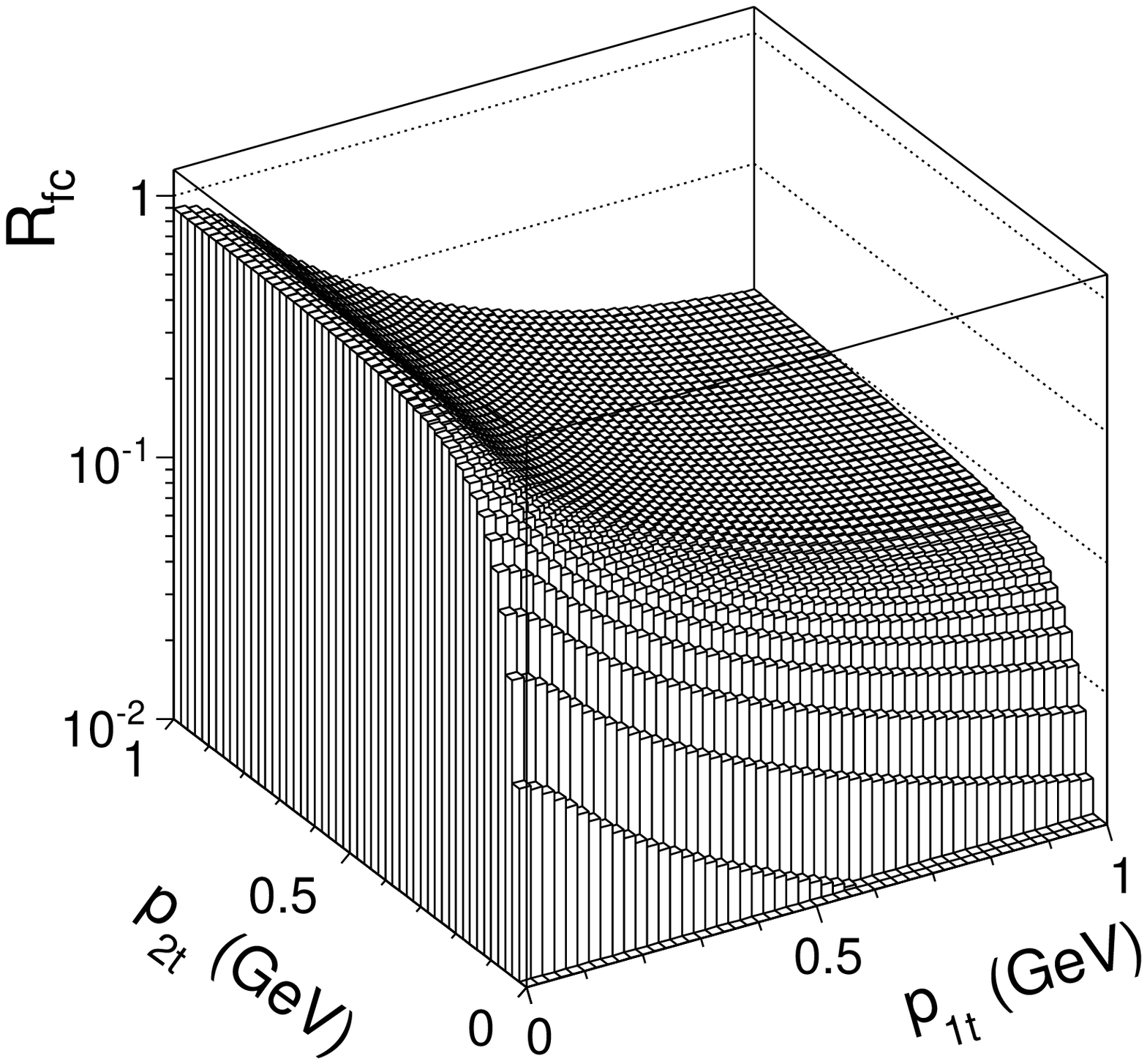}
   \caption{\label{fig:map_pt1pt2_pipi_f0_1500_rat}
   \small Helicity decomposition of the cross section
          on the ($p_{1t},p_{2t}$) plane for W = 10 GeV. 
          $R_{cc}$ (upper left), $R_{ff}$ (upper right),
          $R_{cf}$ (lower left), $R_{fc}$ (lower right).
          The standard nucleon dipole form factor was used in 
          this calculation.
}
\end{figure}

Up to now we have neglected interference between pion-pion and
pomeron-pomeron contributions (for the same final channel). 
This effect may be potentially important when both components are 
of the same order of magnitude.
At J-PARC energies there could be, in principle, some small interference
effect. \footnote{At the PANDA energies the problem is rather academic
as the diffractive component can be neglected.}
While the pomeron-pomeron contribution is dominantly nucleon helicity
preserving the situation for pion-pion fusion is more complicated.
In the latter case we define 4 classes of contributions with respect
to the nucleon helicities: $cc$ (both helicities conserved),
$cf$ (first conserved, second flipped), $fc$ (first flipped, 
second conserved) and $ff$ (both helicities flipped).
The corresponding ratios of individual contributions to the sum of 
all contributions are shown in 
Fig.\ref{fig:map_pt1pt2_pipi_f0_1500_rat}. 
In practice, only the $cc$ $\pi \pi$ contribution may potentially 
interfere with the gluonic one. From the figure one can conclude 
that this can happen only when both transverse momenta of 
the final nucleons are small. We shall leave numerical studies of 
the interference effect for future investigations, when experimental 
details of such measurements will be better known; but already now one 
can expect them to be rather small.


\begin{figure}[!htb]    
\includegraphics[width=0.5\textwidth]{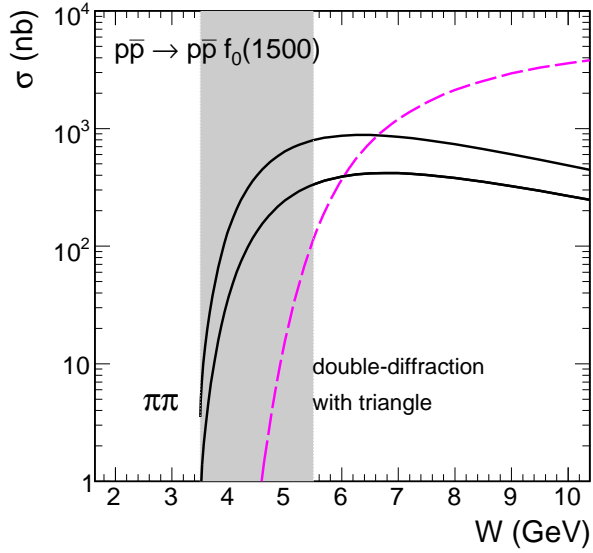}
   \caption{\label{fig:pipifusion_triangle}
   \small  Comparison of the pion-pion fusion component (solid lines) and 
   the double-diffractive one with pionic triangle (dashed line). The details
   concerning the double-diffractive component are explained in section II-D.
   The vertical gray band shows the range of the center-of-mass energy available
   by the PANDA experiment.
}
\end{figure}

Now we wish to show the size of the triangle-double-diffractive (TDD) component
at the GSI HESR energy range.
In Fig.\ref{fig:pipifusion_triangle} we compare it with the pion-pion
fusion component. The TDD component vanishes quickly with decreasing
energy and stays below the pion-pion fusion component for the HESR
energy range. The quick decrease of the cross section is caused mainly
by the $F_{cut}(s_{1,eff})$ and  $F_{cut}(s_{2,eff})$ factors in Eq.(\ref{diffractive_pion_triangle_amplitude})
and (\ref{WpiN_correction_factor})
and reflects smallness of $\pi N$ subchannel energies.

\section{Background for the $f_0(1500)$ production for $p\bar p \to p\bar p f_0(1500)$ reaction}
\subsection{Pion-pion rescattering background}

In the previous section we have shown that in the PANDA
energy range the pion-pion fusion is the dominant reaction
mechanism for the production of the glueball candidate
$f_0(1500)$.
Up to now we have calculated the cross section
for production of $f_0(1500)$ meson -- a process
with three particles ($p$, $\bar p$ and $f_0(1500)$) 
in the final state.

\begin{figure}[!htb]    
 \centerline{\includegraphics[width=0.45\textwidth]{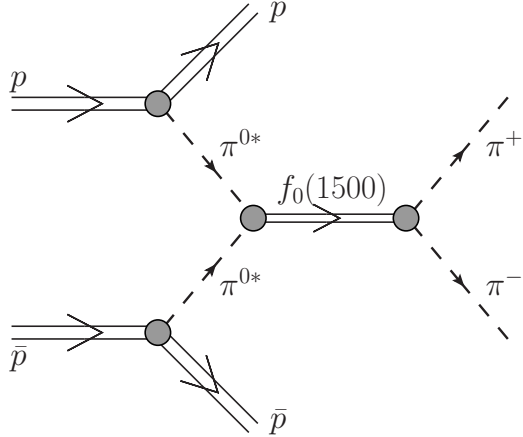}}
   \caption{\label{fig:diagram_2to4_f0_1500}
   \small  The 2 $\to$ 4 amplitude for the reaction
through the glueball candidate $f_0(1500)$. 
The stars attached to $\pi^0$ mesons
denote the fact that they are off-mass-shell.}
\end{figure}

In practice one must select a given decay channel
of $f_0(1500)$.
There are a few options: (a) a two-pion
decay ($\pi^+ \pi^-$ or $\pi^0 \pi^0$), 
(b) a four-pion decay
(c) a two-kaon decay. 
The first one is attractive
due to its simplicity but may have a large background.
The second requires more complicated analysis but may
have smaller background. The branching fraction for 
the last option is smaller by a factor of about 5 than for the
two-pion channel.

\begin{figure}[!htb]    
\includegraphics[width=0.3\textwidth]{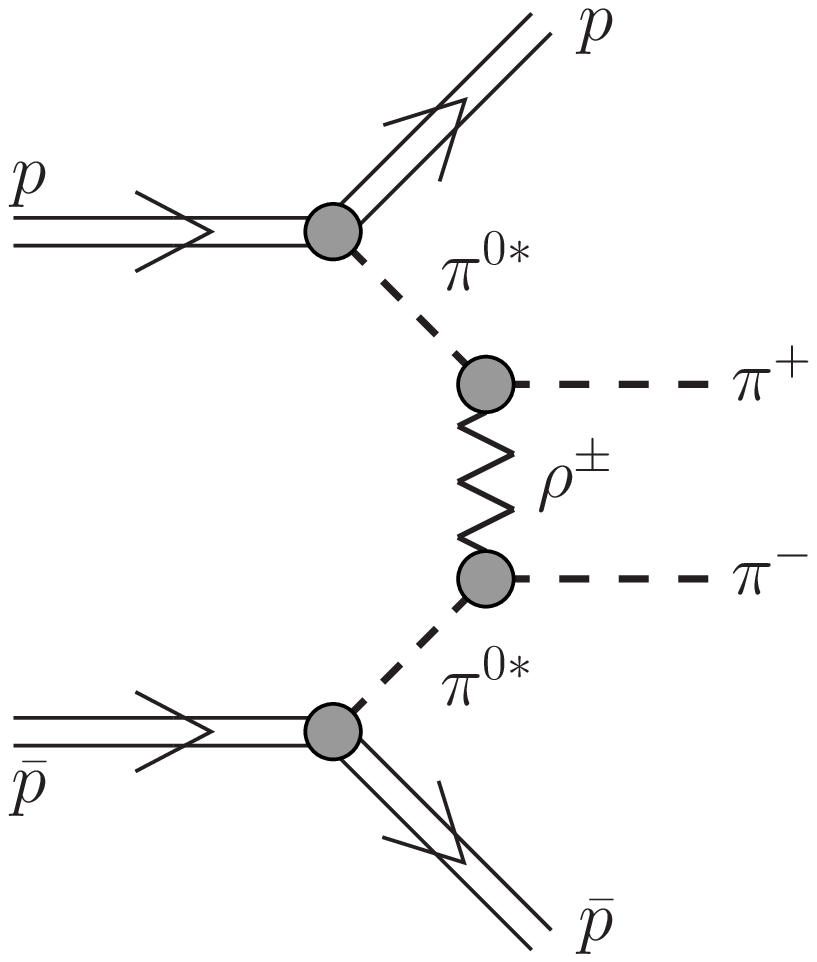}
\space \space \space
\includegraphics[width=0.3\textwidth]{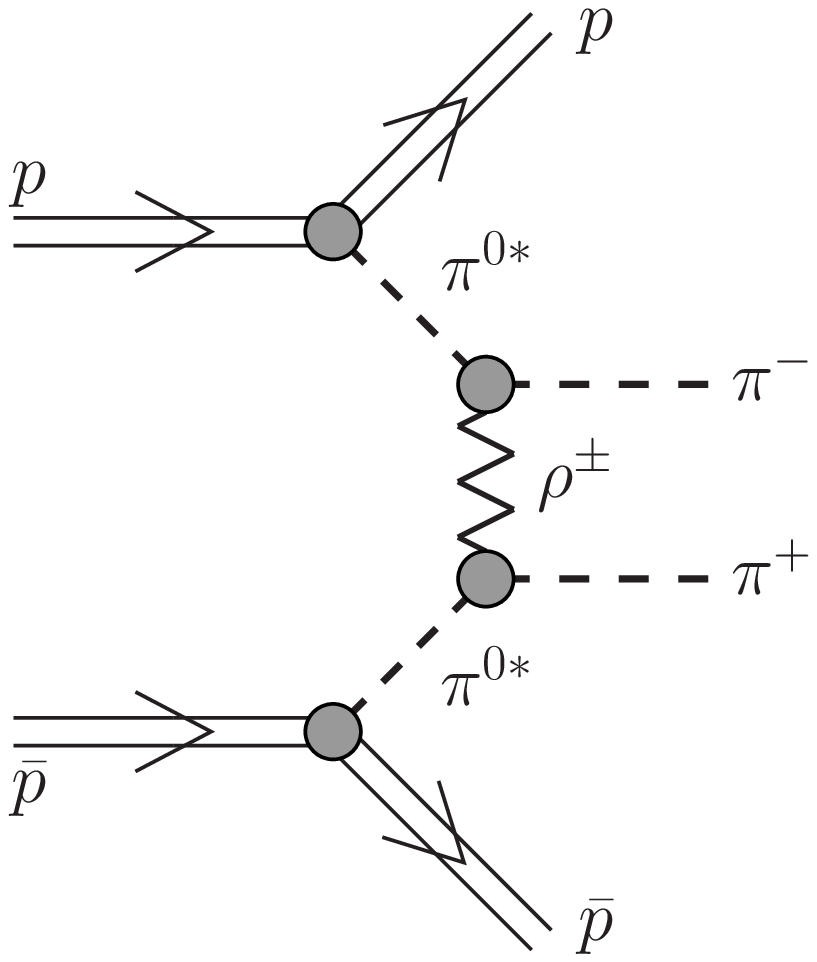}
   \caption{\label{fig:diagram_2to4_background}
   \small  The 2 $\to$ 4 amplitudes with the
intermediate $\rho$-meson (reggeon) exchange
as an example of the background.}
\end{figure}

Let us consider now an estimate of the background
to the $p \bar p \to p \bar p \pi^+ \pi^-$ reaction.
In Fig.\ref{fig:diagram_2to4_f0_1500} we present
our reaction of interest -- the reaction which proceeds
through the scalar resonance $f_0(1500)$. 
This reaction is viewed now as a process with four
particles ($p, \bar p, \pi^+, \pi^-$) in the final state. 
Unavoidably there exists a nonreduceable background to 
this process.
In Fig.\ref{fig:diagram_2to4_background} we show 
an example of the background. We shall call these 
two complex diagrams as $\rho$-meson(reggeon) exchanges 
for brevity.
The region of $W_{\pi \pi} \sim$ 1.5 GeV is slightly 
above the region of application of the standard 
meson-exchange formalism and slightly below the region 
of application of high-energy Regge approach.
In principle, one should consider both approaches.

In the $\rho$ meson-exchange formalism the reduced amplitude
for the $\pi^0 \pi^0 \to \pi^+ \pi^-$ process can
be written as:
\begin{eqnarray}
{\cal M}_{\pi^0 \pi^0 \to \pi^+ \pi^-}^{\rho-exch.} 
&=&
g_{\pi \pi \rho} F_{\pi \pi \rho}({\hat t})
\frac{(q_1^{\mu}+p_3^{\mu}) P_{\mu \nu} (q_2^{\nu}+p_4^{\nu})}
{{\hat t} - m_{\rho}^2 + i \Gamma_{\rho} m_{\rho}}
g_{\pi \pi \rho} F_{\pi \pi \rho}({\hat t})
\nonumber \\
&+&
g_{\pi \pi \rho} F_{\pi \pi \rho}({\hat u})
\frac{(q_1^{\mu}+p_4^{\mu}) P_{\mu \nu} (q_2^{\nu}+p_3^{\nu})}
{{\hat u} - m_{\rho}^2 + i \Gamma_{\rho} m_{\rho}}
g_{\pi \pi \rho} F_{\pi \pi \rho}({\hat u}) \;.
\label{amplitude_rho_exchange}
\end{eqnarray}
Above
\begin{equation}
P_{\mu \nu}(k) = -g_{\mu \nu} +  k_{\mu} k_{\nu} / m_{\rho}^2
\; .
\label{P_munu}  
\end{equation}
The quantities $F_{\pi \pi \rho}(k^2)$ in 
(\ref{amplitude_rho_exchange}) 
describe couplings of extended objects: 
pions and the exchanged $\rho$-meson.
We parameterize them in the exponential form:
\begin{equation}
F_{\pi \pi \rho}(k^2) = \exp
\left(
\frac{k^2 - m_{\rho}^2}{\Lambda^2}
\right)
= \exp\left(
\frac{B_{\pi \pi \rho}}{4}(k^2-m_{\rho}^2)
\right)
 \; .
\label{pipirho_formfactors}
\end{equation}
Consistent with the definition of the coupling constant
the form factors are normalized to unity when $\rho$
meson is on-mass-shell. We take
$\frac{g_{\pi \pi \rho}^2}{4 \pi}$ = 2.6, which 
reproduces the $\rho$ meson decay width \cite{PDG},
and $\Lambda$ = 1 GeV ($B_{\pi \pi \rho}$ = 4 GeV$^{-2}$).

In the case of $\rho$-reggeon exchange the amplitude 
can be written as
\begin{equation}
{\cal M}_{\pi^0 \pi^0 \to \pi^+ \pi^-}^{\rho-reggeon} 
=
s_{34} \eta_{\rho}({\hat t}) C_{\pi \pi \rho} F({\hat t}) 
\left( \frac{s_{34}}{s_0} \right)^
{\alpha_{\rho}({\hat t})-1 } 
F({\hat t}) \\
+ 
s_{34} \eta_{\rho}({\hat u}) C_{\pi \pi \rho} F({\hat u})
\left( \frac{s_{34}}{s_0} \right)^
{\alpha_{\rho}({\hat u})-1 }
F({\hat u})  \; .
\label{amplitude_reggeon_exchange}
\end{equation}
We parameterize the vertex form factors in the standard
exponential form used usually in the Regge 
phenomenology
\begin{equation}
F(k^2) = \exp \left( \frac{B_{\pi \pi \rho}}{4} k^2 \right) \; .
\end{equation}
In practical calculation we take 
$B_{\pi \pi \rho}$ = 6 GeV$^{-2}$.
In the formula (\ref{amplitude_reggeon_exchange}) 
$\eta_{\rho}(k^2)$ is a signature
factor which we take here $\eta_{\rho} \approx i + 1$ 
and $\alpha_{\rho}(t/u) = \alpha_{\rho}(0)
+\alpha_{\rho}' \cdot  t/u$
is a so-called reggeon trajectory.
We take from the phenomenology:
$\alpha_{\rho}(0)$ = 0.5475 \cite{DL92} and 
$\alpha_{\rho}'$ = 0.9 GeV$^{-2}$.
The strength parameter $C_{\pi \pi \rho}$ can be 
obtained assuming Regge factorization
(see e.g.\cite{SNS02}) and using the known strength
parameters for the $N N$ and $\pi N$ scattering
fitted to the corresponding total cross sections 
\cite{DL92}.
The simple Regge parameterizations apply for energies
$W >$ 2 GeV (see e.g.\cite{DL92}). In our case of
pion-pion scattering energies of $W \sim$ 1.5 GeV
are of interest. Here a small modification of the
Regge formula (\ref{amplitude_reggeon_exchange}) 
may be in order.
Consistent with meson-exchange formalism (spin-1 exchange)
one may expect saturation of the 
$\pi^0 \pi^0 \to \pi^+ \pi^-$
cross section at lower energies.
The following freezing of the energy factor
in (\ref{amplitude_reggeon_exchange}) seems a reasonable 
correction:
\begin{equation}
\left(\frac{s_{\pi \pi}}{s_0} \right)^{\alpha_{\rho}}
\to \left(\frac{s_{freez}}{s_0} \right)^{\alpha_{\rho}}
\; ,
\label{freezing_reggeon}
\end{equation}
where $s_{freez} = W_{freez}^2$.
One may expect $W_{freez}$ = 1.5 -- 2.0 GeV. 
The compatibility of the Regge formalism with low-energy
approaches for pion-pion scattering was discussed 
in Ref.\cite{PY04}. 

The 2 $\to$ 2 amplitudes (\ref{amplitude_rho_exchange})
and (\ref{amplitude_reggeon_exchange})
may be inserted into the 2 $\to$ 4 amplitude of 
Fig.\ref{fig:diagram_2to4_background}.
When doing so we include in addition the correction for
off-shellness of incoming pions as was done for
the $f_0(1500)$ meson using exponential form factors (as in \ref{off-shell_form_factors}).
Now we can perform a genuine 2 $\to$ 4 calculation
including four-body phase space.

\begin{figure}[!htb]    
\includegraphics[width=0.5\textwidth]{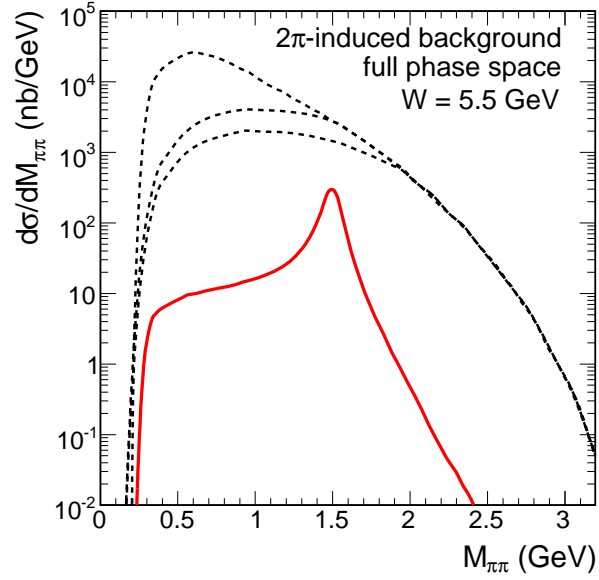}
   \caption{\label{fig:dsigma_dMpipi_full}
   \small  
Pion-pion invariant mass distribution for $f_0(1500)$ (solid line)
and the $\rho$-reggeon exchange background (dashed lines)
for naive (upper line) and corrected (two lower lines with 
$W_{freez}$ = 1.5, 2 GeV) extrapolations to low energies. 
Here the full phase space has been included.
The calculation was performed for the highest PANDA
center-of-mass energy $W$ = 5.5 GeV.}
\end{figure}

Here we discuss the results with $\rho$-reggeon exchange
only. We have checked that the $\rho$-meson exchange
formalism discussed in this section provides 
approximately the same results at $W_{\pi \pi} <$ 1.2 GeV
as the modified reggeon exchange
with $W_{freez}$ = 1.5 -- 2.0 GeV (see formula 
(\ref{freezing_reggeon})). Therefore the modified 
reggeon-exchange calculation provides a realistic 
predictions in the broad range of pion-pion energies,
both above and below the $f_0(1500)$ resonance.

In Fig.\ref{fig:dsigma_dMpipi_full} we show 
two-pion invariant mass distribution. 
The solid line corresponds to our resonance contribution. 
The dashed lines correspond to the 
$\rho$-reggeon exchange contribution. Here the resonance 
contribution is much lower than the $\rho$-exchange 
background.
In this calculation the integration over whole 
phase space was done.

\begin{figure}[!htb]    
\includegraphics[width=0.5\textwidth]{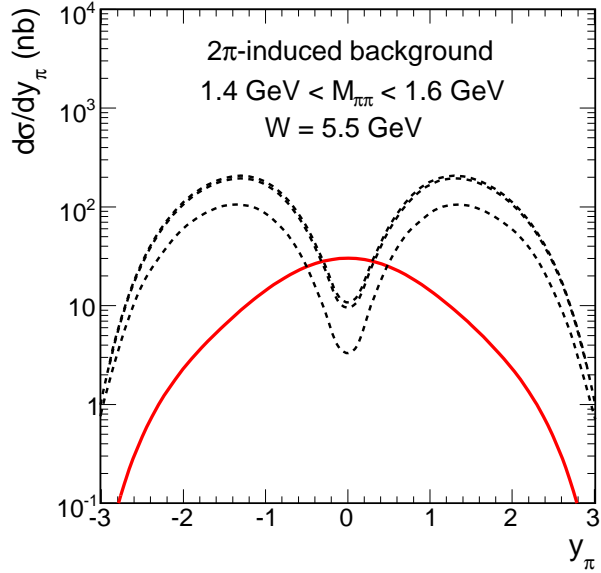}
   \caption{\label{fig:dsigma_dypi}
   \small  
Rapidity distribution of pions from the decay of the glueball candidate
$f_0(1500)$ (solid line) and
from the $\rho$-reggeon exchange background (dashed lines)
for naive (upper line) and corrected (lower lines) extrapolations 
to low energies. For the background we impose in addition:
1.4 GeV $< M_{\pi \pi} <$ 1.6 GeV.
The calculation was performed for the highest PANDA
center-of-mass energy $W$ = 5.5 GeV.}
\end{figure}

\begin{figure}[!htb]    
\includegraphics[width=0.5\textwidth]{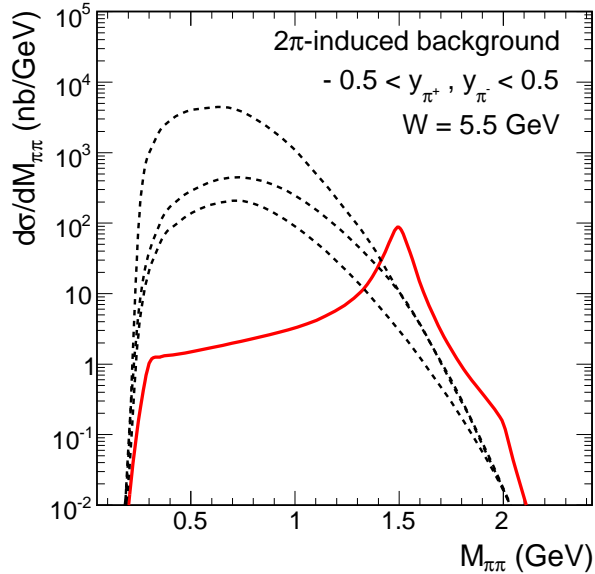}
   \caption{\label{fig:dsigma_dMpipi_cuty3y4}
   \small  
Pion-pion invariant mass distribution for 
$f_0(1500)$ (solid line) and the $\rho$-reggeon exchange 
background (dashed lines)
for naive (upper line) and corrected (lower lines with 
$W_{freez}$ = 1.5, 2.0 GeV) extrapolations to low energies. 
Here an additional condition on center-of-mass rapidities
-0.5 $< y_{\pi^+}, y_{\pi^-} <$ 0.5 has been imposed.
The calculation was performed for the highest PANDA
center-of-mass energy $W$ = 5.5 GeV.}
\end{figure}

In Fig.\ref{fig:dsigma_dypi} we show corresponding 
distributions in rapidity of $\pi^+$ or $\pi^-$ 
(identical). In order to better see the overlap of
the signal and background for the $\rho$-reggeon exchange
we impose in addition 1.4 GeV $< M_{\pi \pi} <$ 1.6 GeV
(the region of the $f_0(1500)$ resonance). 
Limiting to very small center-of-mass rapidities one can
further improve the signal-to-background ratio.
In Fig.\ref{fig:dsigma_dMpipi_cuty3y4} we show
two-pion invariant mass distribution with extra cuts:
-0.5 $< y_{\pi^+}, y_{\pi^-} <$ 0.5. 
While the $f_0(1500)$ contribution is only slightly
modified, the $\rho$-reggeon background contribution
is reduced by more than order of magnitude.
One can clearly see the signal over background in 
this case. Especially the high-energy side of
the $f_0(1500)$ meson is now free of the $\rho$-exchange
background.
A better separation can be done by using pion-pion 
partial wave analysis.

Close to the two-pions production threshold the
Roper resonance excitation and its subsequent decay 
($N^{*}(1440) \to N \pi \pi$) is known to give 
the dominant contribution to 
the $p p \to p p \pi^+ \pi^-$ reaction \cite{AOH98}.
The same may be expected also for the 
$p \bar p \to p \bar p \pi^+ \pi^-$ reaction. 
The Roper resonance produces the two-pions in 
dominantly the $l$=0 and $I$=0 state
\footnote{Here $l$ is angular momentum between
pions and $I$ is the total isospin of the pion pair.}
(the tail of the $\sigma$ meson), 
i.e. the strength is concentrated at $M_{\pi \pi}$ 
much lower than $f_0(1500)$.
The kinematical constraint gives
$M_{\pi \pi} < M_{N^{*}(1440)} - M_{N} \approx$ 0.5 GeV.
In addition, this contribution could be eliminated by 
extra cuts on invariant masses $M(p \pi^+ \pi^-)$
and $M(\bar p \pi^+ \pi^-)$.
The same method can, at least in principle, be used
to eliminate the double $\Delta$, $\bar \Delta$ excitations
followed by their decays
$\Delta \to \pi p$ and $\bar \Delta \to \pi \bar p$
\footnote{We have checked that eliminating the region
of double-Delta excitation at the highest PANDA energy
$W$ = 5.5 GeV reduces the signal by less than about 5\%.}.
In this sense the last two contributions (Roper and 
double isobar excitations) are reduceable.
To which extend precision of the real apparatus will allow 
such a reduction is a matter of further investigations.

Certainly complete analysis requires including more
processes and an analysis of cuts allowing for improving
the signal-to-background ratio. This certainly goes beyond
the scope of the present paper where we only signal 
a huge increase of the $f_0(1500)$ meson production 
cross section in the PANDA energy range
due to pion-pion fusion, the mechanism never
discussed before in the literature.

The $p \bar p \to p \bar p \pi \pi \pi \pi$ reaction
may be more favorable as far as the signal-to-background 
ratio is considered. Unfortunately theoretical calculation
of background are not feasible in this case. 
It is not clear to us at 
present if the 6-body channel can be measured
by the PANDA detector at FAIR.

In the case of the $p \bar p \to p \bar p K^+ K^-$
reaction the relevant branching fraction is smaller,
but not negligible 
($f_0(1500) \to K \bar K$ = 8.6 \% \cite{PDG}).
On the other hand the contribution from nucleon 
resonances is probably considerably smaller. 
There is, however, unreduceable contribution from 
$K^*$ exchange in the $K^0 \bar K^0 \to K^+ K^-$ 
subprocess.
The parameters for the latter reaction are much less
known than those for the $\pi^0 \pi^0 \to \pi^+ \pi^-$
subprocess.

\subsection{Double-diffractive two-pion background}

At high energies another two-pion continuum may be of interest.
\begin{figure}
\begin{center}
  \includegraphics[width=5.5cm]{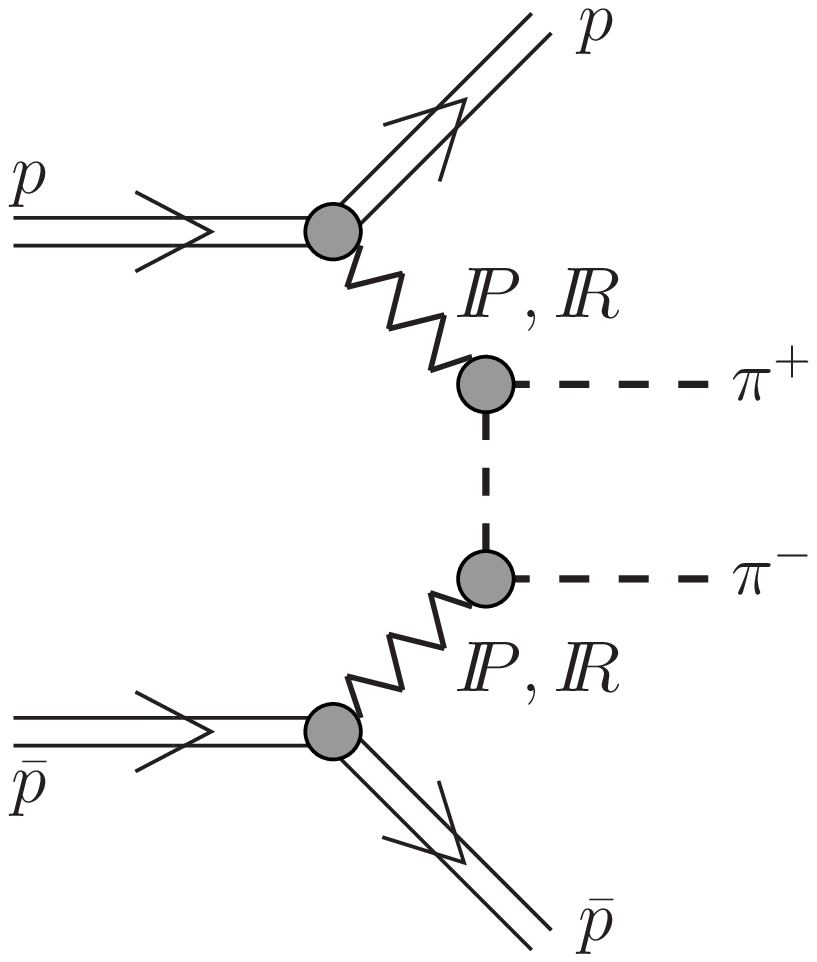}
  \includegraphics[width=5.5cm]{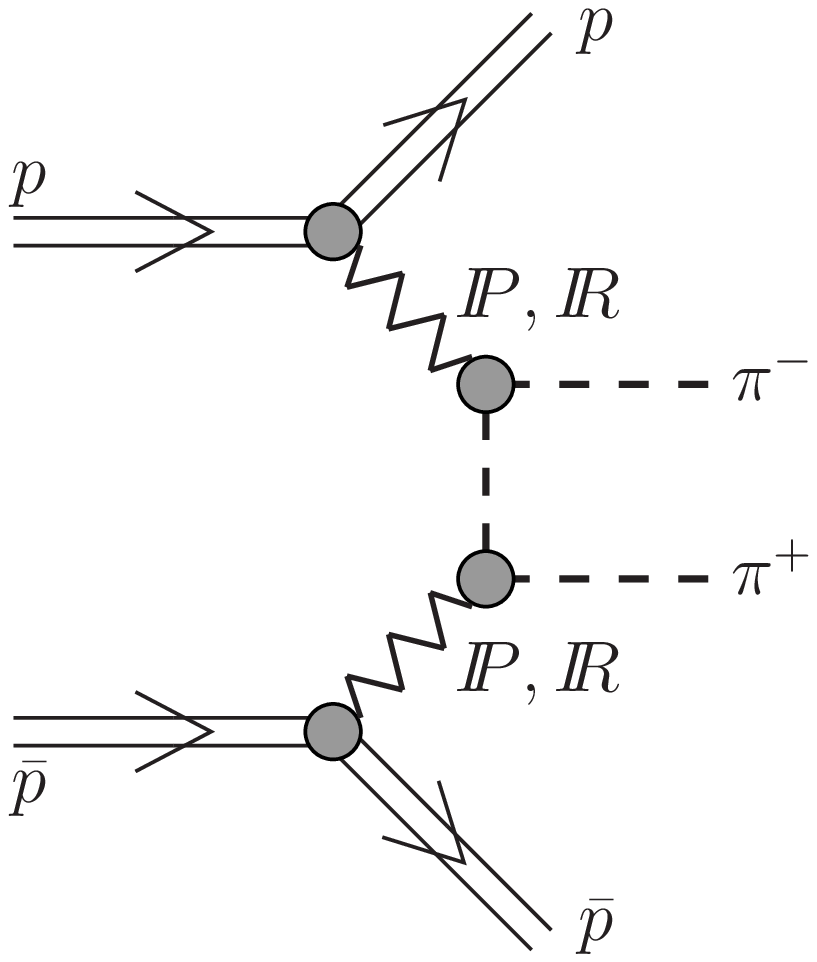}
\end{center}
  \caption{A sketch of the double-diffractive 
mechanisms of exclusive production of 
the $\pi^+ \pi^-$ pairs.
\label{fig:2pi_mechanisms}
}
\end{figure}
The underlying mechanism was proposed long ago in 
Ref.\cite{PH76}. The general situation is sketched 
in Fig.\ref{fig:2pi_mechanisms}.
The corresponding amplitude for the
$p p \to p p \pi^+ \pi^-$ process (with four-momenta 
$p_a + p_b \to p_1 + p_2 + p_3 + p_4$) can be written
as
\begin{eqnarray}
{\cal M}^{p p \to p p \pi^+ \pi^-} &=&
M_{13}(t_1,s_{13}) \; F(t_a) \; 
\frac{1}{t_a - m_{\pi}^2} \;
F(t_a) \; M_{24}(t_2,s_{24}) \nonumber \\
&+&
M_{14}(t_1,s_{14}) \; F(t_b) \;
\frac{1}{t_b - m_{\pi}^2} \;
F(t_b) \; M_{23}(t_2,s_{23})
 \;,
\label{Regge_amplitude}
\end{eqnarray}
where $M_{ik}$ denotes "interaction" between nucleon $i$=1 (forward nucleon) or
$i$=2 (backward nucleon)
and one of the two pions $k$=3 $(\pi^+)$, $k$=4 $(\pi^-)$. 
In the Regge phenomenology they can be written as:
\begin{eqnarray}
M_{13}(t_1,s_{13})&=& is_{13} 
   \left(  C_{I\!\!R}^{13} \left(\frac{s_{13}}{s_0}\right)^{\alpha_{I\!\!R}(t_{1})-1} \; e^{\frac{B_{\pi N}}{2} \; t_1}
        +  C_{I\!\!P} \left(\frac{s_{13}}{s_0}\right)^{\alpha_{I\!\!P}(t_{1})-1} \; e^{\frac{B_{\pi N}}{2} \; t_1}
   \right) \; ,
\nonumber \\
M_{14}(t_1,s_{14})&=& is_{14}
   \left(  C_{I\!\!R}^{14} \left(\frac{s_{14}}{s_0}\right)^{\alpha_{I\!\!R}(t_{1})-1} \; e^{\frac{B_{\pi N}}{2} \; t_1}
        +  C_{I\!\!P} \left(\frac{s_{14}}{s_0}\right)^{\alpha_{I\!\!P}(t_{1})-1} \; e^{\frac{B_{\pi N}}{2} \; t_1}
   \right) \; ,
\nonumber \\
M_{24}(t_2,s_{24})&=& is_{24} 
   \left(  C_{I\!\!R}^{24} \left(\frac{s_{24}}{s_0}\right)^{\alpha_{I\!\!R}(t_{2})-1} \; e^{\frac{B_{\pi N}}{2} \; t_2}
        +  C_{I\!\!P} \left(\frac{s_{24}}{s_0}\right)^{\alpha_{I\!\!P}(t_{2})-1} \; e^{\frac{B_{\pi N}}{2} \; t_2}
   \right) \; ,
\nonumber \\
M_{23}(t_2,s_{23})&=& is_{23}
   \left(  C_{I\!\!R}^{23} \left(\frac{s_{23}}{s_0}\right)^{\alpha_{I\!\!R}(t_{2})-1} \; e^{\frac{B_{\pi N}}{2} \; t_2}
        +  C_{I\!\!P} \left(\frac{s_{23}}{s_0}\right)^{\alpha_{I\!\!P}(t_{2})-1} \; e^{\frac{B_{\pi N}}{2} \; t_2}
   \right) \; .
\label{Regge_propagators}
\end{eqnarray}

Above $s_{ik} = W_{ik}^2$, where $W_{ik}$ is the 
center-of-mass energy in the $(i,k)$ subsystem.
The first terms describe the subleading reggeon exchanges 
while the second terms describe exchange of the 
leading (pomeron) trajectory.
We have neglected real parts of the reggeon 
exchanges amplitudes for simplicity.
The strength parameters of the $\pi N$ interaction
are taken from Ref.\cite{DL92}.

The extra form factors $F(t_{a})$ and $F(t_{b})$ "correct" 
for off-shellness of the intermediate pion
in the middle of the diagrams shown in Fig.\ref{fig:2pi_mechanisms}.
In the fallowing they are parametrized as
\begin{equation} 
F(t)=\exp\left(\frac{t-m_{\pi}^{2}}{\Lambda^{2}_{off}}\right) \;,
\label{off-shell_form_factors}
\end{equation}
i.e. normalized to unity on the pion-mass-shell.
We take $\Lambda_{off}$ = 1 GeV.
More details of the calculation will be presented 
elsewhere \cite{LS09_ddpi}.
The $2 \to 4$ amplitude (\ref{Regge_amplitude}) is used
to calculate the corresponding cross section including
limitations of the four-body phase-space.

\begin{figure}[!htb]    
\includegraphics[width=0.5\textwidth]{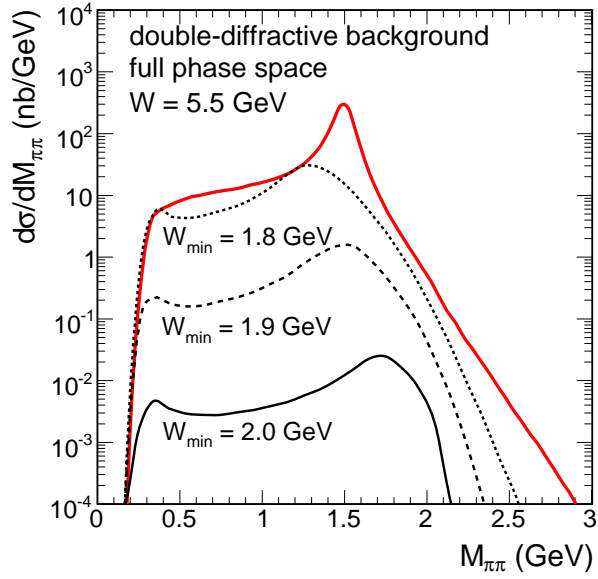}
   \caption{\label{fig:dsigma_dmpipi_full_space_b}
   \small  
Pion-pion invariant mass distribution for $f_0(1500)$ (solid line)
and the double-diffractive background (dashed lines)
for different (sharp) cut-off for $W_{\pi N}$. 
Here the full phase space has been included.
The calculation was performed for the highest PANDA
center-of-mass energy $W$ = 5.5 GeV.}
\end{figure}

To excludes resonance regions we "correct" the Regge parametrization (\ref{Regge_propagators})
by multiplying byfactors $F_{cut}(s_{ik})$ (as in \ref{WpiN_correction_factor}).
In Fig.\ref{fig:dsigma_dmpipi_full_space_b} we show
the two-pion invariant mass distribution of
the double-diffractive background together with
the $f_0(1500)$ signal ($\pi \pi$ fusion).
We show three curves corresponding to different
cuts on both $W_{\pi N}$: $W_{min}$ = 2.0 (solid), 
1.9 (dashed), 1.8 (dotted) GeV.
The figure suggests that the double-diffractive background
should not disturb observing the $f_0(1500)$ signal
at the PANDA experiment energies.

\begin{figure}[!htb]    
\includegraphics[width=0.5\textwidth]{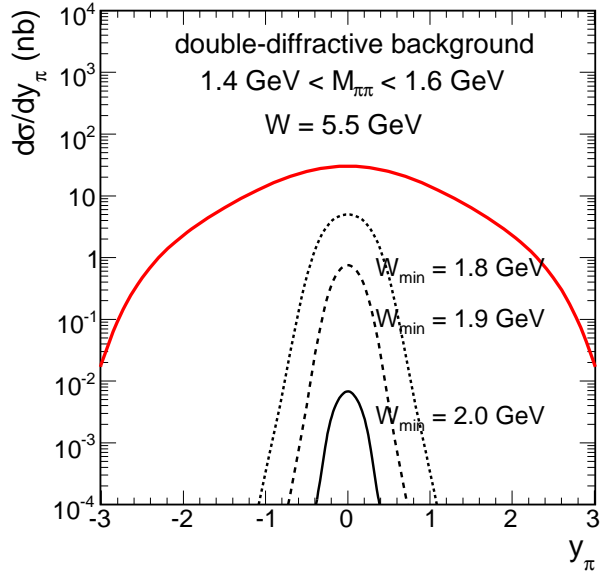}
   \caption{\label{fig:dsigma_dypi_b}
   \small  
Rapidity distribution of pions from the decay of the glueball candidate
$f_0(1500)$ (solid line) and
from the double-diffractive background (dashed lines)
for different (sharp) cut-off for $W_{\pi N}$. 
For the background we impose in addition:
1.4 GeV $< M_{\pi \pi} <$ 1.6 GeV.
The calculation was performed for the highest PANDA
center-of-mass energy $W$ = 5.5 GeV.}
\end{figure}

In this case, unlike for the two-pion involved $\rho$ 
exchange discussed in the previous subsection, 
imposing cuts on pion rapidities would not
be helpful as the double-diffractive contribution
is concentrated at midrapidities as shown
in Fig.\ref{fig:dsigma_dypi_b}.
At lower HESR energies the situation is better,
the double-diffractive two-pion background is  
relatively smaller. 

\section{Discussion and Conclusions}

For the first time in the literature we have estimated the cross 
section for exclusive $f_0(1500)$ meson (glueball candidate) production 
not far from the threshold.
We have included both gluon induced diffractive and triangle-double-diffractive
mechanisms as well as the pion-pion exchange contributions. 

The QCD diffractive component was obtained by extrapolating down the cross section
in the Khoze-Martin-Ryskin approach with unintegrated gluon distributions 
from the literature as well as using two-gluon impact factor approach. 
A rather large uncertainties are associated with 
the QCD diffractive component.
At present only upper limit can be obtained for the diffractive
component as the $f_0(1500) \to g g$ decay coupling constant remains 
unknown. The coupling constant could be
extracted only in high-energy exclusive production of $f_0(1500)$
where other mechanisms are negligible.

We have found rather large contribution of pionic-triangle-double-diffractive component
at higher energies ($ W >$ 10 GeV). However, at the GSI HESR energies
this contribution is strongly damped because of the phase space
limitations on the $\pi N$ subchannel energies. 

A future experiment at RHIC could contribute to shed some light
on the competition of the both diffractive mechanisms.

The calculation of the MEC contribution requires
introducing extra vertex form factors.
At largest PANDA energies they are relatively well known and
the pion-pion fusion can be reliably calculated.
The situation becomes more complicated very close to the threshold
where rather large $|t_1|$ and $|t_2|$ are involved.
The cross section for energies close to the threshold is very sensitive
to the functional form and parameters of vertex form factor.
Therefore a measurement of $f_0(1500)$ close to its production 
threshold could limit the so-called $\pi N N$ form factors 
in the region of exchanged four-momenta never tested before. 

We predict the dominance of the pion-pion contribution
close to the threshold. Our calculation shows that the diffractive
components (in fact its upper limit for the QCD mechanism) are by more than order of magnitude
smaller than the pion-pion fusion component in the energy region of future
PANDA experiment.

The diffractive components may dominate over the pion-pion component
only for center-of-mass energies $W >$ 15 GeV.
Taking into account rather large uncertainties the predictions
of this component should be taken with some grain of salt.
Clearly an experimental program is required to disentangle 
the reaction mechanism at energies $W >$ 15 GeV.

Disentangling the mechanism of the exclusive $f_0(1500)$ production
not far from the meson production threshold would require study of
the $p \bar p \to p \bar p f_0(1500)$,
$p \bar p \to n \bar n f_0(1500)$ processes with the PANDA detector 
at FAIR and $p p \to p p f_0(1500)$ reaction at J-PARC.
In the case the pion exchange 
mechanism is a dominant process one expects:
$\sigma(p \bar p \to n \bar n f_0(1500)) = 4 \times
\sigma(p \bar p \to p \bar p f_0(1500))$.
At high energies, when the gluonic, or double-diffractive
with intermediate triangle, components dominate over MEC components
$\sigma(p \bar p \to p \bar p f_0(1500)) >
\sigma(p \bar p \to n \bar n f_0(1500))$.

At intermediate energies one cannot exclude a priori subleading 
reggeon exchanges like $\rho \rho$ for instance. 
However, we do not know how to reliably calculate them from first 
principles.
We believe that the distortions from the pion-pion
at low energies and/or distortions from the QCD gluonic 
mechanism at high energy may tell us more and allow 
for a phenomenological analysis taking into account 
the $\rho \rho$ component explicitly.
We leave this problem for a future analysis when experimental
data will be available.

Only a careful studies of different final channels in the broad
range of energies could help to shed light on coupling of (nonperturbative) 
gluons to $f_0(1500)$ and therefore would give a new hint on its 
nature.
The experimental studies of exclusive production of $f_0(1500)$ are 
not easy at all as in the $\pi \pi$ decay channel one
expects a large continuum. We have performed an involved calculation
of the four-body $p \bar p \pi^+ \pi^-$ background.
Our calculation shows that imposing extra cuts should allow to extract
the signal of the glueball $f_0(1500)$ candidate at the highest PANDA 
energy.
A partial wave $\pi \pi$ analysis should be helpful in this context. 
The two-pion continuum will be studied in more detail in our future work.
A smaller continuum may be expected in the $K \bar K$ or 
four-pion $f_0(1500)$ decay channel. This requires, however, 
a good geometrical (full solid angle) coverage and high registration 
efficiencies.
PANDA detector seems to fulfill these requirements, but planning
real experiment requires a dedicated Monte Carlo simulation
of the apparatus.

It is a central problem of our field if $f_0(1500)$ is 
a $q \bar q$ or glueball type.
Unfortunately, our analysis does not 
allow to give a definite answer to this important question.
Some information on baryon-baryon correlation may be helpful 
but certainly not decisive.

If the cross section at high energies (where the contribution
of subleading reggeon exchanges may be neglected) is much
smaller than predicted based on the KMR method
it means that gluons only weakly couple to $f_0(1500)$.
This could provide some indirect information on the 
$f_0(1500)$ structure. A direct comparison of the shape
of differential distributions at high energies
may provide a valuable test of the KMR method originally
proposed for exclusive Higgs production
(the latter experiment is very difficult as very small
statistics is predicted).
A possible disagreement with the prediction for 
exclusive $f_0(1500)$ production at high energies could 
put into question the KMR approach, at present
state of art in the field. Experiments at RHIC could be useful
in this context and could shed light on the nonperturbative coupling 
of gluons to $f_0(1500)$.

\vskip 0.5cm

{\bf Acknowledgements} We are indebted to Roman Pasechnik, 
Wolfgang Sch\"afer, Oleg Teryaev, Leonard Le\'sniak and Andrew Kirk for 
a discussion and Tomasz Pietrycki for a help in preparing 
some diagrams.


\end{document}